\documentclass[a4paper,12pt,useAMS,usenatbib,revtex4]{mn2e}
\usepackage{lineno}
\usepackage{graphicx}
\usepackage{longtable}
\usepackage{amssymb}
\usepackage{tabularx, blindtext}
\usepackage{lscape}
\usepackage{url}

\title[Limb darkening and exoplanets]
{Limb darkening and exoplanets: testing stellar model atmospheres and identifying biases in transit parameters}
\author[Espinoza \& Jord\'an ]{
N\'estor Espinoza$^{1,2}$\thanks{E-mail:nespino@astro.puc.cl}, 
Andr\'es Jord\'an$^{2,1}$\thanks{E-mail: ajordan@astro.puc.cl}\\ 
$^{1}$Millennium Institute of Astrophysics, Vicu\~na Mackenna 4860, Santiago, Chile\\
$^{2}$Instituto de Astrof\'isica, Pontificia Universidad Cat\'olica de Chile, Vicu\~na Mackenna 
4860, Santiago, Chile}
\begin{document}

\date{}

\pagerange{\pageref{firstpage}--\pageref{lastpage}} \pubyear{2015}

\maketitle

\label{firstpage}

\begin{abstract}
  Limb-darkening is fundamental in determining transit lightcurve
  shapes, and is typically modelled by a variety of laws that
  parametrize the intensity profile of the star that is being
  transited. Confronted with a transit lightcurve, some authors fix
  the parameters of these laws, the so-called limb-darkening
  coefficients (LDCs), while others prefer to let them float in the
  lightcurve fitting procedure. Which of these is the best strategy, however, 
  is still unclear, as well as how and by how much each of these can bias
  the retrieved transit parameters. In this work we attempt to clarify those points by
  first re-calculating these LDCs, comparing them to measured
  values from {\em Kepler} transit lightcurves using an algorithm that takes into account uncertainties
  in both the geometry of the transit and the parameters of the
  stellar host. We show there are significant departures from
  predicted model values, suggesting that our understanding of 
  limb-darkening still needs to improve. Then, we show through
  simulations that if one uses the quadratic limb-darkening law to 
  parametrize limb-darkening, fixing and fitting the LDCs can 
  lead to significant biases -up to $\sim 3\%$ and $\sim 1\%$ in $R_p/R_*$, 
  respectively-, which are important for several confirmed and candidate exoplanets. 
  We conclude that, in this case, the best approach is to let the LDCs be free in 
  the fitting procedure. Strategies to avoid biases in data from present and future 
  missions involving high precision measurements of transit parameters are described.
\end{abstract}



\section{Introduction}

It has been known since the first observations of our Sun's surface
that its observed intensity decreases towards the limb. This effect,
termed limb darkening, crucially affects the shape of the transit
signature a planet imprints in the observed stellar flux when passing
in front of its host star.
In practice, limb-darkening is parametrized by laws which depend
on $\mu=\cos(\theta)$, where $\theta$ is the angle between the line of
sight and the normal to a given point of the stellar surface. Some of
the most widely used limb darkening laws in exoplanet transit
lightcurve fitting are given by

\begin{eqnarray*}
\frac{I(\mu)}{I(1)} &=& 1-a(1-\mu)\ \ \ \ \ \ \ \ \ \ \ \ \ \ \ \ \ \ \ \ \ \ \ \ \ \ \textnormal{(the linear law),}\\
\frac{I(\mu)}{I(1)} &=& 1-u_1(1-\mu)-u_2(1-\mu)^2\ \ \ \ \ \ \ \textnormal{(quadratic law),}\\
\frac{I(\mu)}{I(1)} &=& 1-\sum_{n=1}^{4}c_n(1-\mu^{n/2})\ \ \ \ \ \ \ \ \ \ \ \ \ \ \ \textnormal{(non-linear law).}\\
\end{eqnarray*}

\noindent The non-linear law has as a special case the square-root law
proposed by \cite{diazgimenez1992}, which is obtained by setting
$c_3=c_4=0$, and the variant, three-parameter law, introduced by
\cite{sing2009}, with $c_1=0$.

The laws listed above do not include all the parametrizations
available for limb-darkening. \cite{klinglesmith1970}, for example,
introduced the logarithmic law for early type stars, which is given by
\begin{eqnarray*}
\frac{I(\mu)}{I(1)} &=& 1-l_1(1-\mu)-l_2\mu \ln \mu,
\end{eqnarray*}
while \cite{claret2003} introduced an exponential law given by
\begin{eqnarray*}
\frac{I(\mu)}{I(1)} &=& 1-e_1(1-\mu)-e_2/(1-e^\mu), 
\end{eqnarray*}
but these laws are less used for transit fitting in practice, mainly
due to their complex forms which are harder to deal with
computationally and are not implemented in the most widely used
transit modelling codes to date \citep[][see, however,
\cite{kjurkchieva2013,abu2013}]{ma2002,eastman2013}.

Confronted with measurements of an exoplanetary transit, observers
usually model the effects of limb darkening by either fitting the
limb-darkening coefficients (LDCs) of a given law or fixing some (or all) of
them using tabulated values \citep[see e.g.,][for some recent tables
using the {\em Kepler} bandpass]{claret11,sing2010}. However, exactly
which strategy is optimal, and in which situations, is still
unclear. Previous works \citep[e.g., ][]{csizmadia2013, muller2013}
have discussed this issue focusing on the uncertainty introduced on
the parameters retrieved from transit observations but, to our
knowledge, no study has yet addressed the potential bias introduced by
them. Such study is called for as such biases could be limiting (1) the
instruments currently obtaining high precision photometry
like {\em Kepler}, (2) exoplanet population studies which by definition are based on averaging 
out random errors but not systematic ones \citep[e.g.,][]{schlaufman2015} and/or (3) techniques that require high-precision
measurements like transmission spectroscopy. Studying the biases
introduced by the treatment of limb darkening and whether or not they
are significant is the main aim of this work.

The sources of potential biases introduced by assuming a given
limb-darkening law can be separated in four. One issue is the
differences in the tabulated values of the limb-darkening
coefficients, which are large even when the same model stellar
atmospheres are used. For example, \cite{csizmadia2013} has shown that the
differences between different approaches at tabulating limb-darkening
coefficients for the quadratic law with the ATLAS9 \citep{kurucz1979}
stellar atmosphere models can be as high as $20\%$, which if
incorporated in the modelling can lead to significant increases in
the uncertainties of the retrieved transit parameters.  The second
issue is the fact that, as shown by \cite{howarth2011}, limb-darkening
coefficients obtained directly from the intensity profiles cannot be
compared directly to coefficients observed in transit photometry due
to the fact that the optimization procedures are different in each
setting. This in turn implies that using LDCs
obtained from an intensity profile of the star as inputs in transit
photometry should lead to biases in the retrieved transit
parameters. The third issue is related to model complexity: we usually
model limb-darkening, which in models is best represented by the
non-linear law, with low-order laws such as the quadratic
law. Finally, the fourth issue, and perhaps the most complicated to
tackle, is the fact that we still do not know with certainty if
available models do a good job at reproducing real intensity profiles
of stars.

The paper is organized as follows. In \S2 we detail our methodology
for obtaining LDCs from
ATLAS9 \citep{kurucz1979} and PHOENIX \citep{Husser2013} stellar model
atmospheres, and compare our results with the literature in order to
try understand the discrepancy between previously published tables. In
\S3 we compare our model LDCs to {\em Kepler}
estimates for a sample of stars, taking in consideration the effects
mentioned by \cite{howarth2011}, and compare the performance of the
ATLAS and PHOENIX models at predicting the observed limb-darkening
effect. In \S4 we explore the biases introduced on different transit
parameters by fixing or fitting the LDCs, \S5 presents a discussion of 
our results and \S6 the conclusions of our work.


\section{Fitting limb-darkening models}

The fitting of the limb-darkening laws to intensity profiles obtained
from stellar model atmospheres is, in principle, a relatively
straightforward procedure. Given a normalized response function for a
given telescope/detector $S_\lambda(\lambda)$ (e.g., $\int
S_\lambda(\lambda)d\lambda=1$, although the normalization doesn't
really matter in practice for the calculation of limb-darkening
coefficients), one must integrate the specific intensity
$I_{\lambda}(\lambda, \mu)$ at each angle $\mu_i=\cos(\theta_i)$
multiplied by the response function, i.e.,

\begin{eqnarray}
\label{specific}
I(\mu_i)=\int_{\lambda_1}^{\lambda_2} I_{\lambda}(\lambda,\mu_i)S_\lambda(\lambda)d\lambda,
\end{eqnarray}

\noindent where $\lambda_1$ and $\lambda_2$ define the wavelength
limits of the band. This gives the observed (by the instrument)
intensity, which can then be fitted by any of the laws cited earlier
after normalizing by $I(1)$. Of course, for CCDs, the recorded
quantity are photons and therefore we have to divide 
the integrand in equation~(\ref{specific}) by $hc/\lambda$, where $h$
is Planck's constant and $c$ is the speed of light.

In this work we make use of two widely used model intensity libraries
to calculate LDCs: the ATLAS9 model
atmospheres, available from Robert L.\ Kurucz's
webpage\footnote{\url{http://kurucz.harvard.edu/grids.html}} and the
1D PHOENIX model atmospheres \citep{Husser2013}. These models differ
both in the geometry used to solve the stellar atmosphere (with ATLAS
models using a plane-parallel approximation and the PHOENIX models
using a spherically symmetric atmosphere), and on the actual physics
used on each of them \citep[see, e.g.,][for a short up-to-date
discussion on this matter]{plez2011}, so a difference between the
LDCs computed from both models is expected.  We
will derive and compare LDCs from these stellar
atmosphere models using the {\em Kepler} high-resolution response
function\footnote{\url{http://keplergo.arc.nasa.gov/kepler_response_hires1.txt}}
in order to compare our results with those in the literature.

\subsection{Fitting limb-darkening laws with the ATLAS models}

\begin{figure*}
\includegraphics{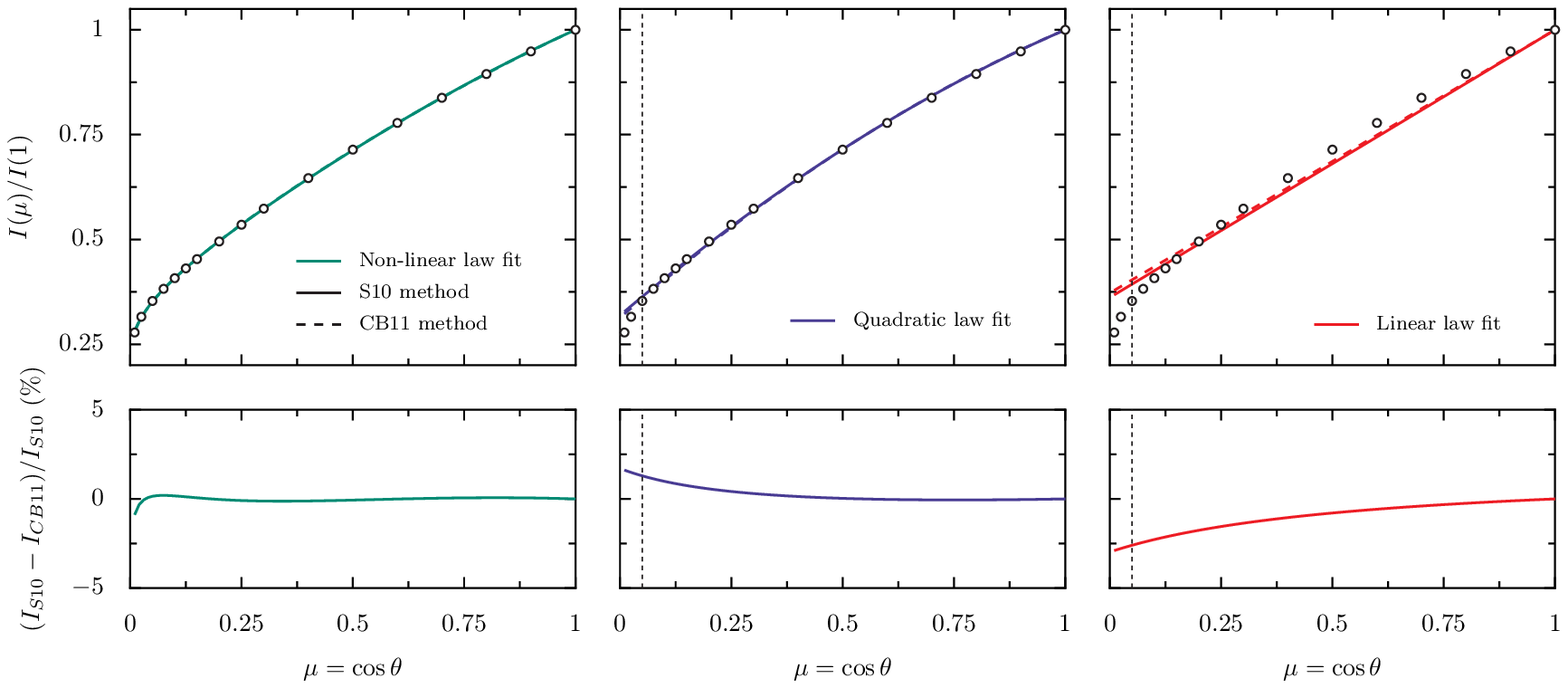}
\caption{ATLAS stellar intensity profile obtained for the {\em Kepler} bandpass for a G5V type star with solar
  metallicity (white points), with the colored lines showing fits to the most popular limb-darkening laws. In 
  the upper panel, solid lines correspond to fits obtained following the method of S10,  while the dashed lines 
  correspond to fits obtained by following the method of CB11 (note these overlap in the 
  leftmost panels). The lower panel shows the (percentual) difference 
  between these fits. The vertical dashed black line marks $\mu=0.05$ (see text).}
  \label{intensity_atlas}
\end{figure*}

We first note that the ATLAS intensities are given per unit frequency
and not per unit wavelength, so a a $c/\lambda^2$ term has to be added
in Equation~(\ref{specific}), and so the integral in this equation 
now reads

\begin{eqnarray*}
I(\mu_i) = \int^{\lambda_2}_{\lambda_1} \left(I_\nu(\lambda,\mu_i)\frac{c}{\lambda^2} \frac{\lambda}{hc}\right)S_{\lambda}(\lambda)d\lambda,
\end{eqnarray*}

\noindent where the limits of integration, $\lambda_1=348$ nm and
$\lambda_2=970$ nm are, in our case, the first and last wavelengths
present on the {\em Kepler} 
response function. We recall, however, that in order to fit the laws
cited in the introduction what we want is the intensity normalized  by 
$I(1)$, i.e.,

\begin{eqnarray}
\label{normint}
\frac{I(\mu_i)}{I(1)} = \frac{\int^{\lambda_2}_{\lambda_1} I_\nu(\lambda,\mu_i)S_{\lambda}(\lambda)/\lambda\ d\lambda}{\int^{\lambda_2}_{\lambda_1} 
I_\nu(\lambda,1)S_{\lambda}(\lambda)/\lambda\ d\lambda}. 
\end{eqnarray}

After using numerical integration and interpolation to 
perform these integrals, the resulting normalized intensity profiles were fitted by a least-squares 
procedure using two different approaches in order to compare our results with previous
works. The first approach was to fit all the angles for the
non-linear law, while fitting only intensities at $\mu\geq 0.05$ for
the rest of the laws, a procedure followed by \cite{sing2010},
hereafter referred to as S10. The second approach was to fit to
100-$\mu$ points obtained by interpolating the 17 angles given by the
ATLAS models\footnote{At first sight, fitting to 100 points that have 
been interpolated from a set of only 17 points given by the models 
seems to lack any justification. We show in what follows that there 
is a rationale for this procedure and that there is a well defined 
criterion under which this procedure gives better results.} in a
linear grid from $\mu=0.01$ to $\mu=1.0$ in $0.01$ steps using a cubic
spline, 
a procedure followed by \cite{claret11}, hereafter refered as
CB11\footnote{We will use S10 and CB11 to refer both to the papers and
  the methodologies assumed in those works to determine the
  LDCs}.

We note that, because the coefficients are linear with respect to the
parameters in all of the laws considered, the least-squares solutions
are unique and therefore have analytic solutions given a set of
intensities and angles. The exact solution, outlined in Appendix~A,
was used to perform the least-square fits. Figure~\ref{intensity_atlas} shows the intensity profiles
for a G5V star of solar metallicity (i.e., with an effective
temperature of $T_\textnormal{eff}=5500$, log-gravity of $\log g=4.5$,
$[M/H]=0.0$ and microturbulent velocity of $v_\textnormal{turb}=2$
km/s) where, for illustration, fits to the most popular limb-darkening
laws used in the literature have been plotted following the methods of
S10 (solid lines) and CB11 (dashed lines).
The lower panel shows the (percentual) difference between those two
methods, illustrating that they actually differ by a small (but, as we
will show, significant) amount, with the differences being larger for
low order laws such as the linear (with a median difference of $\sim
1\%$, reaching a $\sim 2.5$\% difference near the limb) and smaller
for high-order laws like the non-linear (with a median difference of
$\sim 0.1\%$, and reaching a $\sim 1\%$ difference near the extreme
limb). As one would expect, the difference between the methods of S10
and CB11 is more important for the low order laws.

\subsubsection{Comparing LDCs to previous results}

\begin{figure}
\includegraphics{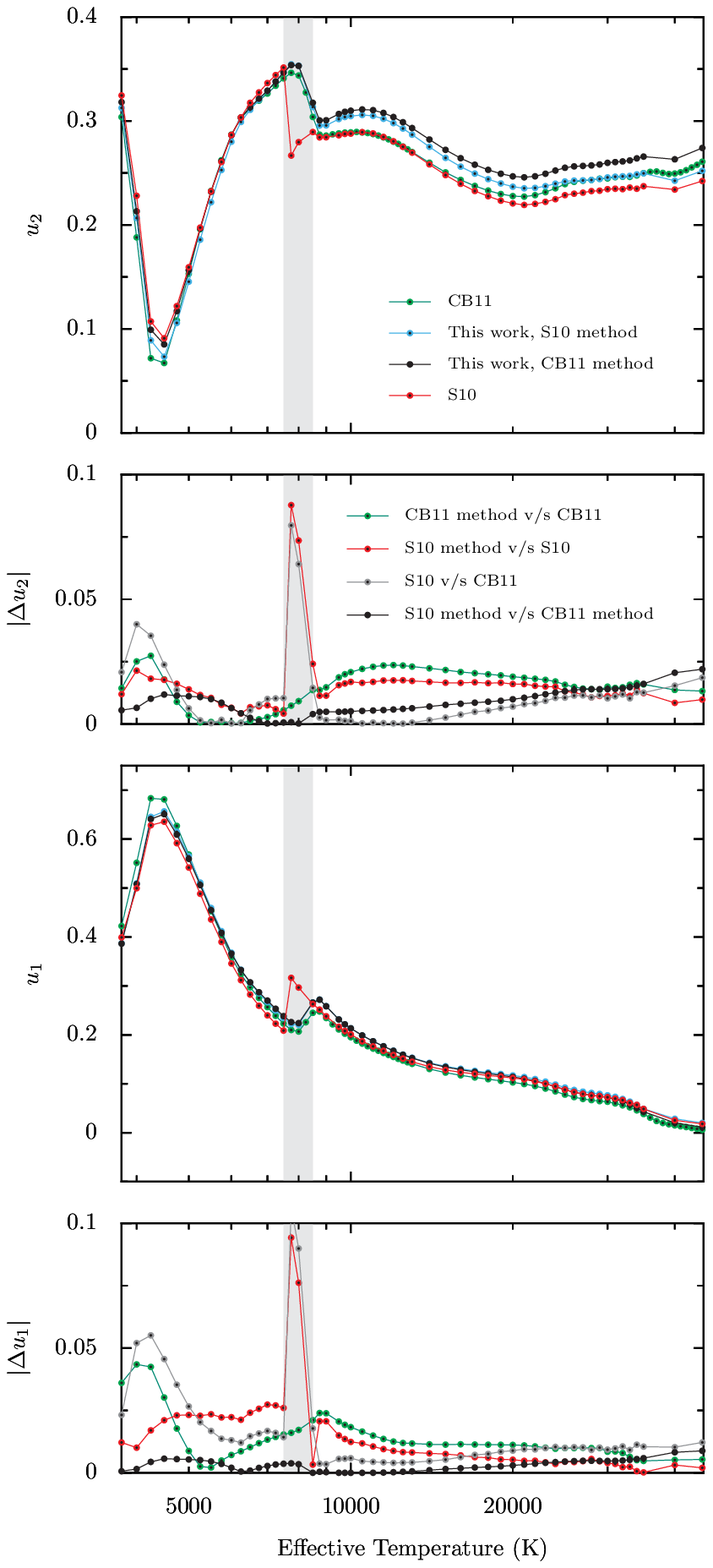}
\caption{Limb-darkening coefficients for the quadratic law obtained
  through the methods described in this work (blue and black), along
  with previous results by Sing (2010, red) and Claret \& Bloemen
  (2011, green). Below each graph the absolute difference
  between our results and each previous study is shown, while
  the difference between those studies is depicted by gray points and
  between the S10 and CB11 methods (but with our procedures) in black
  points. Note the large differences in the range
  $T_\textnormal{eff}\approx 7500-8500$ (region denoted by the gray
  bands) with the Sing (2010) results; this is due to Sing (2010)
  using old versions of the ATLAS stellar model atmospheres.}
  \label{lds_kepler}
\end{figure}

The large panels in Figure~\ref{lds_kepler} show the limb-darkening
coefficients $u_1$ and $u_2$ for the quadratic law, for a range of
effective temperatures, $T_\textnormal{eff}$, calculated  by us using the ATLAS models
for stars with solar metallicity ($[M/H]=0.0$), $\log g = 4.5$ and
$v_\textnormal{turb}=2$ km/s using the S10 and CB11 procedures
(blue and black points), along with the same calculations made by
S10\footnote{The coefficients plotted here can be found in David
  Sing's webpage: \url{http://www.astro.ex.ac.uk/people/sing/David_Sing/Limb_Darkening.html}. In
  particular, we used the full version of Table 3.} (red points), and
CB11 (green points). The small panels just below each large one depict the 
absolute difference between our values and the ones
obtained by those previous studies.

Overall, it can be seen that the coefficients we calculate follow
similar trends as the ones presented in previous studies, except for
the region between $T_\textnormal{eff}\approx 7500-8500$ (gray bands
in Figure~\ref{lds_kepler}) where both our results and the ones of
CB11 show a smooth decrease in the $u_1$ coefficient and a smooth
increase in the $u_2$ coefficient, while the S10 results show a sharp
increase and decrease in $u_1$ and $u_2$ respectively. We found that
this is due to S10 using old versions of the ATLAS stellar atmospheres\footnote{See 
the note on {http://kurucz.harvard.edu/grids.html}; the new models are the ones
  ending in *new.pck (e.g., \texttt{im01k2new.pck}). The old ones
  (e.g., \texttt{im01k2.pck19}) are also available on the webpage
  (Kurucz, private comm.).}.
Using those old models (available in
Robert L. Kurucz's webpage) we recover the overall shape of the
coefficients presented by S10 but cannot, however, eliminate the
differences between S10 and the LDCs we
calculate using the S10 methods.

Despite the similarity of the overall trend, there are significant
differences between the LDCs we calculate and
previous studies, even though we use their methods as described in
their works and the same stellar atmospheres.
The differences are of the same order of magnitude ($\sim 10\%$) for
both coefficients of the quadratic law, with larger differences in the
$u_2$ coefficient for cooler stars ($T_\textnormal{eff}\lesssim 5000$
K), and in the $u_1$ coefficient for hotter stars
($T_\textnormal{eff}\gtrsim 30000$ K). Around solar temperatures
($5000\ \textnormal{K} \lesssim T_\textnormal{eff}\lesssim 6000$ K)
the differences are smaller, $\sim 0.1-1\%$ between our results and
CB11, and $\sim 1-10\%$ between our results and S10. Although we show 
here only the differences with the quadratic LDCs, there are very significant 
deviations between the coefficients of other limb-darkening laws too. In 
particular, we note that the limb-darkening 
coefficients of the non-linear law obtained by the different methods vary 
widely between works and, thus, these must be used with caution. 
We rule out that the differences arise from different interpolation
and/or integration methods by performing the same calculations in
different ways and in different programming environments. 

Given the results above, it is clear that the differences with
previous studies do not arise from different numerical approaches but,
rather, they arise either from the actual method used to obtain the
LDCs (i.e., from the very definition of the
integrals used to estimate the integrated intensities) or from
differences in the model atmospheres used to obtain those coefficients
(e.g., older versions of the ATLAS model atmospheres). Unveiling the
actual source of these discrepancies is currently not feasible as
the actual codes used to obtain these tables are not publicly
available. We believe that in the era of high precision measurements
making all details available for scrutiny in order to try understand
discrepancies is fundamental. Following this logic, we provide all 
the codes needed to reproduce the results in this paper through
GitHub\footnote{\texttt{http://www.github.com/nespinoza/limb-darkening}}. 
Our codes can be used to obtain LDCs for arbitrary
response functions.

\subsubsection{Comparison of the methods used to fit the limb-darkening laws}
\label{sec:lim}

We now discuss which of the two fitting methods, i.e., that of S10 or that 
of CB11, is the most appropriate for obtaining LDCs using stellar 
model atmospheres, as they clearly show significant differences in
Figure~\ref{lds_kepler} (black points on the small panels). 


To assess which of the two methods is better, we define below a
criterion that relies on an accurate analytic description of the
intensity profile given by the ATLAS models.
To this end, we follow \cite{howarth2011} and use the non-linear
limb-darkening law as an accurate descriptor of the intensity profile
for the same stellar parameters used in Figure~\ref{lds_kepler}. We
note that this was done not to try to exactly reproduce the model
intensity profiles but, rather, to emulate an intensity profile that
is similar to the one that would be actually observed in a real
stellar atmosphere. From this
profile, we then obtain quadratic LDCs by
sampling $N=17$ points at the same $\mu$ values as the ones given by
the ATLAS models and, with this, fit the intensity profiles using the
methods of CB11 and a method similar to that of S10 (for this
experiment, we choose to fit all the $\mu$ values and not only the
values for which $\mu \geq 0.05$, which is the original method of S10,
in order to have a fair comparison between the coefficients obtained
by these two methods).

\begin{figure}
\includegraphics{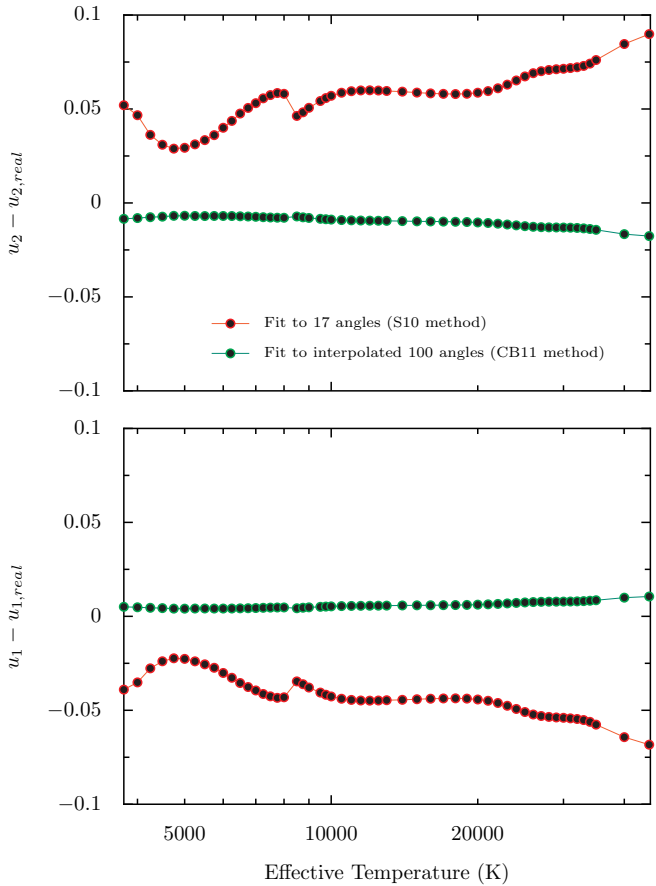}
\caption{Differences between LDCs for the quadratic law obtained
  following the different methods discussed in this work (S10 in red,
  CB11 in green) and the ``real'' underlying quadratic
  LDCs.}
  \label{lds_kepler_methods}
\end{figure}

In order to assess how good the retrieved coefficients are, we
obtain the ``real'' underlying quadratic LDCs
by defining them as the coefficients one would recover when sampling a
very large number of points $N$ from the (real) profile and then
fitting those points with the quadratic limb-darkening law. Under this
definition, when $N\to \infty$, a set of ``limiting coefficients''
should be recovered, which are the ``real'', underlying quadratic
LDCs that best-fit the underlying profile. We
thus take the limit as $N\to \infty$ in our least-squares procedure,
imposing that the data are sampled from an underlying profile
described by the non-linear law (see Appendix B). The resulting
limiting coefficients for the quadratic law are given in terms of the
coefficients of the underlying non-linear law by

\begin{eqnarray}
u_{1,\textnormal{real}}=\lim_{N\to \infty}{u_1} &=& \frac{12}{35}c_1 + c_2 + \frac{164}{105}c_3 +2c_4, \nonumber\\
u_{2,\textnormal{real}}=\lim_{N\to \infty}{u_2} &=& \frac{10}{21}c_1 -
\frac{34}{63}c_3 - c_4.
\label{eq:limcoeffs}
\end{eqnarray}

\noindent Figure \ref{lds_kepler_methods} shows the results of this
experiment.
The fitting method of CB11 is clearly better at obtaining the ``real''
quadratic LDCs as we have defined them. The
reason is that a cubic interpolation of the profile does a good job at
retrieving the underlying profile (the non-linear law in the case of
our experiment) and, thus, sampling points from this interpolation is
very similar to sampling more points from the profile, and thus a
result closer to the ``real'' coefficients 
follows. It is also interesting to see that both methods fail at
retrieving the ``real'' $u_1$ coefficients for very hot stars. This is
due to the fact that limb darkening for hotter stars is sharper than
for cooler stars, changing abruptly towards the limb.  Because the 17
angles given by the ATLAS models are more densely sampled close to the
limb, this gives more weight to that region of the profile and thus
the fit is dominated by it. Sampling uniformly across the profile,
which is what the method of CB11 does, alleviates this problem and
gives an overall better fit to the {\em whole} profile.

As a final note on this topic, we would like to note that the
limiting coefficients we have introduced are the best by construction because to
obtain them we use a sampling scheme that weights the whole profile
uniformly. To obtain the limiting coefficients one has to make use of
the non-linear LDCs, which are obtained by
fitting the model intensity profiles using one of the discussed
methods. However, as shown in the lower panels of
Figure~\ref{intensity_atlas}, in the case of the non-linear law the
fitting method assumed is not very relevant, producing median offsets
on the order of only $0.1\%$. Because of this, we believe that the ``best''
quadratic LDCs that one can choose are the
``limiting coefficients'', whose input non-linear limb-darkening
coefficients might be obtained from either of the S10 and CB11
methods, or some other similar sampling scheme. 
For practical applications, we recommend using the CB11 method to fit
the non-linear law to a given intensity profile and then use the
resulting coefficients to obtain the limiting coefficients according
to equation~(\ref{eq:limcoeffs}). The codes we provide are able to carry
out these calculations.

\subsection{Fitting limb-darkening laws to the PHOENIX models}

After performing the analysis for the ATLAS stellar atmospheres, we
now describe how we obtain LDCs with the
PHOENIX models.  The method we used is very similar to the one
described for the ATLAS models. except for the fact that now we have
78 angles and, unlike ATLAS models, the intensities are given per unit
wavelength. There is an additional, very important difference: unlike
previous studies, we are careful in taking in consideration that
the geometry of the PHOENIX models is different from that of the ATLAS
models and, thus, the original intensity profiles obtained from these
two model atmospheres {\em are not directly comparable.}

Spherically symmetric models like the PHOENIX ones extend the
atmosphere by a small but important fraction \citep[$\simeq 0.4 \%$
for older versions of PHOENIX,][]{ALK2005} from what one usually calls
the ``stellar radius'', which is where the Rosseland mean optical
depth, $\tau_R$, satisfies $\tau_R\sim 1$. Plane-parallel model
atmospheres like ATLAS, on the other hand, by definition have a very
thin extended atmosphere outside the stellar radius and, therefore,
have $\tau_R \sim 1$ effectively at (or very close to) the limb
($\mu=0$). If we recall that the normalized stellar radius is given by
$r=\sqrt{1-\mu^2}$, this implies that for plane-parallel model
atmospheres this outer ``Rosseland radius'' is effectively at $r=1$
(i.e., at $\mu=0$), while for spherically symmetric models this radius
is at $r<1$ (i.e., at $\mu > 0$). This fact is evident when plotting
the profiles of plane-parallel and spherically symmetric model
atmospheres, where the PHOENIX models have a sudden decrease in intensity
around $\mu \sim 0.05$ that is not present in the ATLAS
profiles. This, of course, is not a fundamental difference between the
models but, rather, a difference arising from the fact that these two
models are defining in a different way what the ``outer stellar 
radius'' or the ``limb'' is.

One way of accounting for the different treatment of the stellar limb
in spherical models is to search for the point $r_\textnormal{max}$ at
which $\tau_R\sim 1$ and re-define that point as having $\mu=0$
\textit{before} fitting any law to the intensity profile. This
procedure allows to have meaningful and comparable profiles between
plane-parallel and spherically symmetric models. According to
empirical results with the PHOENIX models, the point
$r_\textnormal{max}$ in the intensity profile can be found by
searching the value at which the derivative of the intensity profile
with respect to the radial intensity profile, $|dI/dr|$, is maximum
\citep[see, e.g.,][]{WAK2004}. Once found, it suffices then to
re-normalize the radial profile by dividing by $r_\textnormal{max}$,
thus re-defining then the point at which $r=1$ ($\mu=0$).

\begin{figure}
\includegraphics{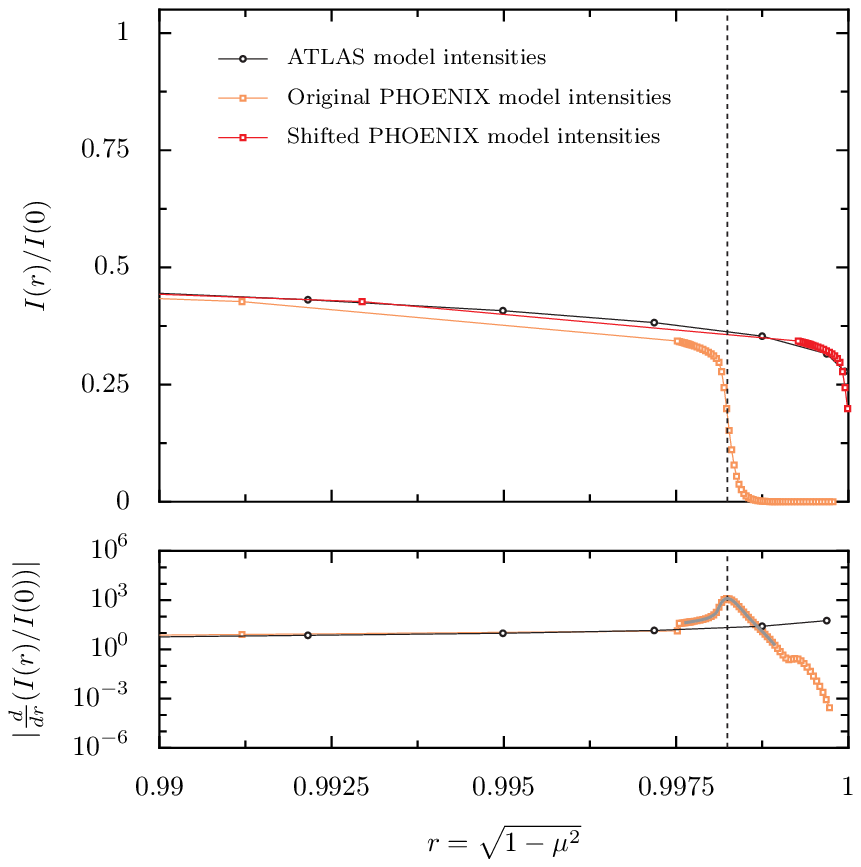}
\caption{Close-up to the original (orange squares) and shifted (red
  squares) limb-darkening profiles for the PHOENIX model atmosphere of
  a G5V type star with solar metallicity (top) and the derivative of
  the intensity profile as a function of $r=\sqrt{1-\mu^2}$
  (bottom). The black circles show the ATLAS profile for
  comparison. The dashed line marks the value of
  $r_\textnormal{max}$ found in this case.}
  \label{lds_kepler_phoenix2}
\end{figure}

Figure~\ref{lds_kepler_phoenix2} shows the results of
applying such corrections to the PHOENIX profiles for a 
G5V type star, where we have plotted the
original PHOENIX profile, the ``shifted'' version of it and the ATLAS
intensity profile for comparison, which also illustrates the sudden 
decrease in intensity already mentioned for the PHOENIX models near $\mu = 0.05$ 
($r\approx 0.9987$). The lower panel of
Figure~\ref{lds_kepler_phoenix2} shows, for illustration, the
derivative of $I(r)/I(0)$, where the maximum is indicated by the dashed
line. It is evident from the upper panel of Figure
\ref{lds_kepler_phoenix2} that the (shifted) PHOENIX and ATLAS
profiles agree for this star at the limb (which was expected because
now both models are modelling the same portions of the stellar
disk). It is also important to see the significant changes between the
original PHOENIX profile and the corrected ones: it is
clear that the original PHOENIX models were not modelling the same
portions of the stellar disk as the plane-parallel models and, thus,
any LDCs derived from it couldn't be directly
compared to the ATLAS models. We note that this correction was not made 
by \cite{claret12,claret13}, hereafter referred as CHW, when deriving their 
``quasi-spherical'' LDCs which were obtained in order 
to be compared to ATLAS models, nor by \cite{NeilsonLester13a,NeilsonLester13b}, 
who also make direct comparisons between the spherical
version of the ATLAS models and their plane-parallel counterparts.

\begin{figure*}
\includegraphics{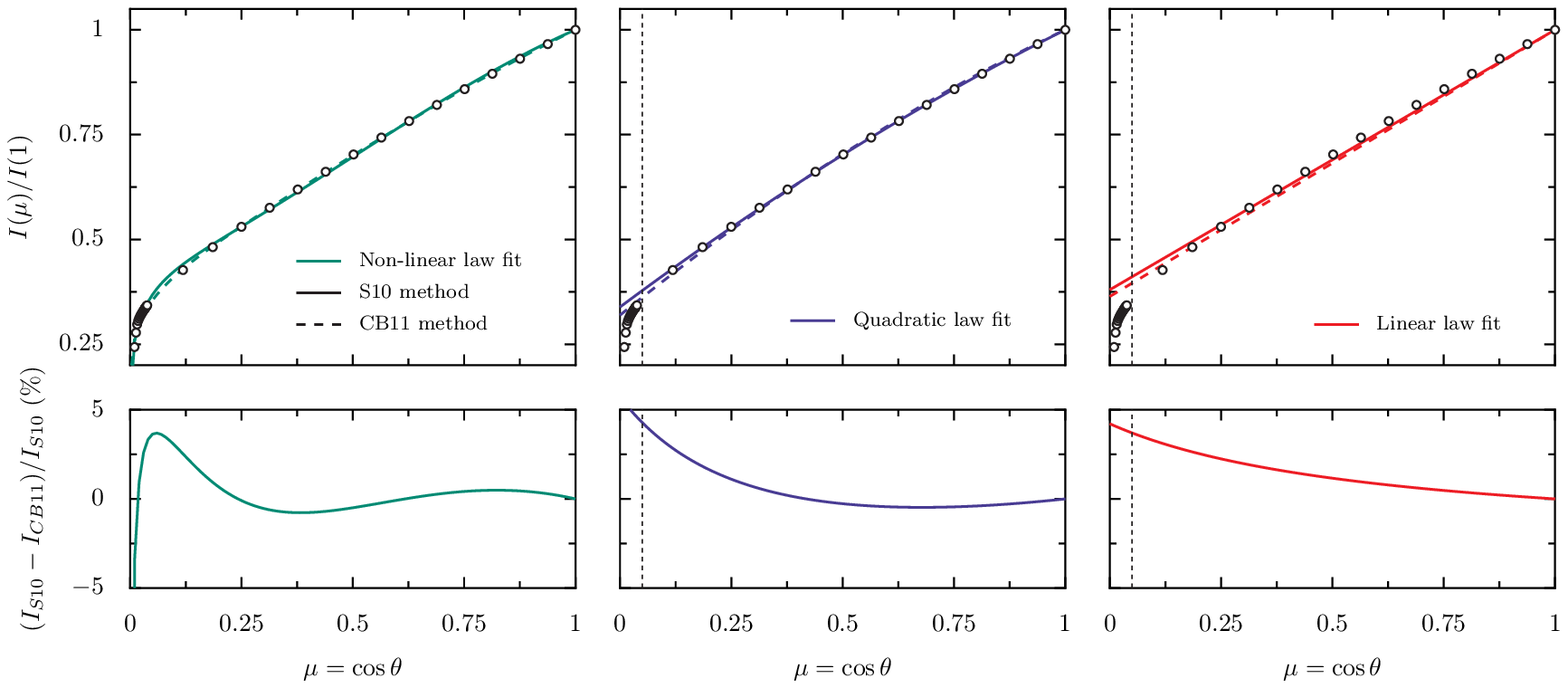}
\caption{PHOENIX stellar intensity profile obtained for the {\em Kepler} bandpass for a G5V type star with solar
  metallicity (white points). The panels show the same information as Figure \ref{intensity_atlas}, but for the ``shifted'' PHOENIX models.}
  \label{lds_kepler_phoenix4}
\end{figure*}

Figure~\ref{lds_kepler_phoenix4} shows fits to the shifted PHOENIX
intensity profiles to the most popular limb-darkening laws used in the
literature for the same model shown in Figure 
\ref{lds_kepler_phoenix2}, following the methods of S10 (solid
lines) and CB11 (dashed lines). Overall, it can be seen that in the
case of the PHOENIX models a larger deviation is observed between the
methods used to fit the profile than for the ATLAS models (compare the
lower panels between this and Figure~\ref{intensity_atlas}). In terms
of following the profile, the CB11 method does a better job than the
S10 method at following the whole profile in the case of the
non-linear law. This was expected in light of our discussion in
Section 2.1.2. As for the quadratic and linear law fits, both methods
seem to do an adequate job at describing the profile given the low
flexibility of the models.

\subsubsection{Comparing LDCs with previous results}

\begin{figure}
\includegraphics{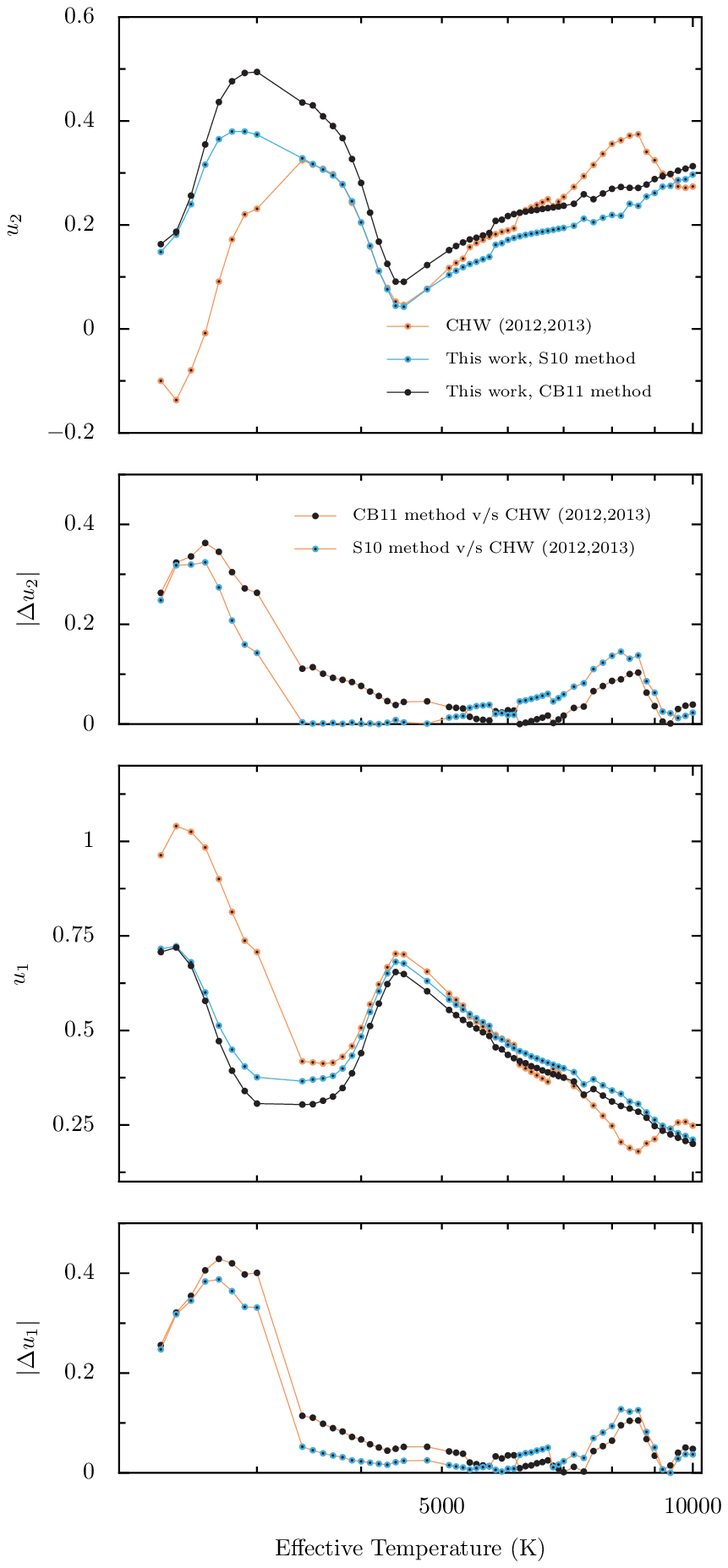}
\caption{Limb-darkening coefficients for the quadratic law obtained
  through the methods described in this work (blue and black), along
  with previous results by Claret, Hauschildt \& Wittle (2012,2013) in
  orange (here we plot the coefficients obtained with the
  ``quasi-spherical models'', which only fit values of $\mu\geq 0.1$
  using the original PHOENIX intensity profiles); below each graph the
  absolute difference between our results and this previous
  study is shown.}
\label{lds_quad_kepler_phoenix}
\end{figure}

Figure~\ref{lds_quad_kepler_phoenix} shows the analogue of
Figure~\ref{lds_kepler} for the quadratic LDCs
obtained using the PHOENIX model atmospheres and the different methods
discussed in this work, where a solar metallicity, $\log g = 4.5$ and
$v_\textnormal{turb}=2$ km/s has been used. For comparison, the
coefficients of CHW have been plotted in orange in the big
panels\footnote{We note that in the models that contain the intensity
  profiles made available to the public by the PHOENIX team
  (\url{http://phoenix.astro.physik.uni-goettingen.de/}) there is no
  stellar model atmosphere for $T_\textnormal{eff}=5000$ K, solar
  metallicity and $\log g =4.5$ and, thus, we have omitted this value
  published by CHW in this plot.}. We can see in the Figure that the
overall trend for stars hotter than $\sim 4000$ K is similar for all
cases except in the region between $\sim 7500-9000$ K, where the
differences are on the order of $\sim 50\%$ and for stars cooler than
$\sim 4000$ K, where the differences are larger than $100\%$. The
overall differences in the coefficients are large in comparison to the
ones seen for the ATLAS models, with median differences on the order
of $\sim 20\%$ for both coefficients with previous works. This is
mainly due to our approach using a new parametrization of the 
stellar disk for obtaining limb-darkening coefficients from spherical 
model atmospheres such as PHOENIX.


\section{Measuring the limb-darkening effect from transit lightcurves}

A quantitative determination of the limb-darkening effect from transit
photometry requires careful thought and has been the subject of some
scrutiny in recent years. For example, \cite{csizmadia2013} has shown
that the presence of unnoculted spots can lead to significant changes
in the LDCs, due to the fact that a spotted
stars' intensity distribution is more complex than the simple laws
given in the introduction. Furthermore, even in the case of unspotted
stars the interpretation of LDCs is
subtle. \cite{howarth2011} showed that the geometry of the transit
biases the LDCs obtained from photometry. In
particular, he shows that high-impact parameter transits highly bias
the observed LDCs because they
 sample chords of the stellar surface which are closer to the
limb and, thus, sample a very different part of the intensity profile
as the one sampled when fitting a whole model intensity
profile. This is an unavoidable problem and, therefore, LDCs obtained 
from high-impact parameter transits are not
directly comparable to the ones obtained by model intensities.

\cite{howarth2011} also stresses that, because the optimization
processes are different when fitting an intensity profile directly
from model stellar atmospheres than when fitting transit lightcurves,
the coefficients obtained by those two procedures are not directly
comparable {\em even if the transit is central}. This is a very
important and often overlooked fact: it implies that limb-darkening
coefficients obtained from transit lightcurves such as the ones
obtained by \cite{muller2013} should not be compared directly to
LDCs obtained from intensity profiles derived
from model stellar atmospheres. That this is relevant can be verified in a
straightforward fashion with a simple simulation study:

\begin{enumerate}

\item Select a good representation of a model intensity profile for a
  given set of stellar parameters derived from stellar model
  atmospheres, such as the non-linear limb-darkening law (i.e., select
  a set of coefficients $c_1$, $c_2$, $c_3$, $c_4$).

\item Generate a (noiseless) synthetic transit lightcurve by feeding
  the chosen representation of the model intensity profile of the star
  and using any set of geometric parameters for the transit.

\item Fit this synthetic transit lightcurve with the same code as the
  one used to generate it, but now using a quadratic limb-darkening
  law parametrization, fixing all the geometric parameters of the
  transit to its input values (i.e., letting just the limb-darkening
  coefficients to float).

\end{enumerate}

\begin{figure}
\includegraphics{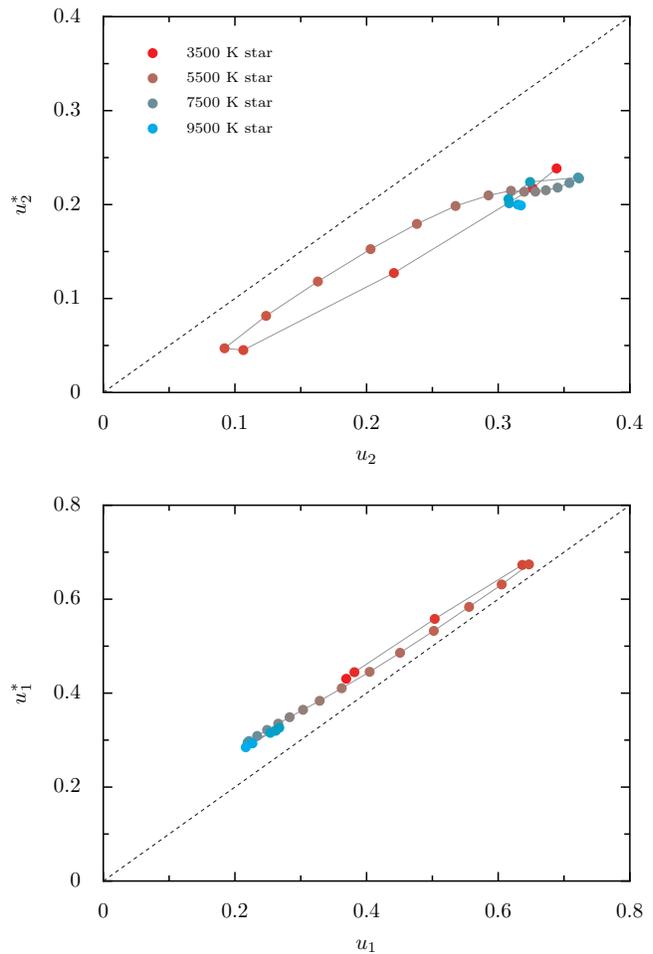}
\caption{Simulation verifying the results of Howarth (2011), where the
  quadratic LDCs obtained by fitting the intensity profiles from model 
  stellar atmospheres ($u_1,u_2$) for stars with $\log g=4.5$, $[M/H]=0.0$ 
  and $v_\textnormal{turb}=2$ km/s using the ATLAS models for different 
  temperatures are plotted against these same coefficients ($u_1^*,u_2^*$) 
  obtained from synthetic transit light-curves generated using these same 
  intensity profiles. The gray line follows the temperatures of the simulated 
  systems in increasing order for better visualization. The dashed
  lines depict the $u_i = u_i^*$ lines, which the points should follow
  if the coefficients from model intensity profiles and transit
  lightcurves were directly comparable.}
\label{simulation_howarth}
\end{figure}

The result of this experiment is always the same and, at first,
counter-intuitive: the quadratic LDCs obtained
from the input model intensity profiles (obtained through, e.g., the 
limiting coefficients obtained directly from the non-linear coefficients 
using equation \ref{eq:limcoeffs}), denoted in what follows as
$(u_1,u_2)$, are \textit{always} different from the ones obtained from
the fit to the synthetic lightcurve, which we denote as
$(u_1^*,u_2^*)$. Figure~\ref{simulation_howarth} shows the results for
this simple experiment, where we used a simulated transit lightcurve
of a planet with a period of $P=3$ days, semi-major axis to stellar
radius ratio $a/R_* = 0.1$, planet-to-star radius ratio $R_p/R_*=0.1$
on a circular ($e=0$, $\omega =0 $), edge-on (inclination $i=\pi/2$)
orbit (a typical ``hot-Jupiter''). The non-linear
law LDCs ($c_1$, $c_2$, $c_3$ and $c_4$) were
obtained for stars with $\log g=4.5$, $[M/H]=0.0$ and
$v_\textnormal{turb}=2$ km/s for different temperatures of interest for 
exoplanet studies using the
ATLAS models via the method of CB11 and the {\em Kepler} response
function, and from these we derived the limiting coefficients
$(u_1,u_2)$. The synthetic transit lightcurves were generated using
the formalism of \cite{ma2002}, and the coefficients $(u_1^*,u_2^*)$
obtained via non-linear least-squares using the Levenberg-Marquardt
algorithm. 1000 points between phases $-0.05$ and $0.05$ were
generated in each simulated lightcurve.

From the experiment we can see that, qualitatively, the effect of
observing the LDCs from photometry is to
overestimate the ``underlying'' $u_1$ coefficient, while the effect
for the $u_2$ coefficient is to underestimate it. The exact mapping
$(u_1,u_2)\to(u_1^*,u_2^*)$, however, is non-trivial and, thus, the
comparison between LDCs obtained from transit
photometry, ($u_1^*$, $u_2^*$), and from intensity profiles obtained
from model stellar atmospheres, ($u_1$, $u_2$), is also
non-trivial. Furthermore, this mapping is in theory also dependent on
the transit parameters, because the optimization process used to fit a
transit lightcurve is dependent on them, although we find this
dependence to be usually negligible.  There is thus no simple rule to
compare LDCs obtained from transit photometry
with the ones obtained by model intensity profiles, and the comparison
has to be done on a case-by-case basis.

\subsection{Comparing photometrically obtained limb-darkening
  coefficients to model values}

In order to compare the LDCs obtained from
photometric measurements to values obtained from model intensity
profiles, \cite{howarth2011} introduced the SPAM
(\textbf{S}ynthetic-\textbf{p}hotometry/\textbf{a}tmosphere-\textbf{m}odel)
algorithm, which can be summarised in three steps:

\begin{enumerate}

\item Fit a transit lightcurve and obtain the best-fit geometric
  parameters $\vec{\theta}_p = \{a/R_*, R_p/R_*, e, \omega, i\}$ and
  best-fit quadratic LDCs
  $\vec{u}^f=\{u^f_1,u^f_2\}$. Obtain the stellar parameters that best
  represent the stellar host (e.g., via spectroscopic observations)
  $\vec{\theta}_* = \{T_\textnormal{eff}, \log g, [M/H],
  v_\textnormal{turb}\}$.
 
\item Generate a synthetic transit lightcurve using the best-fit
  geometrical parameters $\vec{\theta}_p$, and using an accurate
  representation of the model intensity profile of a star with stellar
  parameters $\vec{\theta}_*$, which we choose to be the non-linear
  law representation. 

\item Fit the synthetic transit lightcurve to obtain estimates of the
  geometric parameters and LDCs, but now using
  a quadratic limb-darkening law to parametrize the limb-darkening
  effect. With this, obtain the coefficients
  $\vec{u}^*=\{u_1^*,u_2^*\}$ which can now be directly compared to
  the best-fit coefficients obtained from photometry
  $\vec{u}^f=\{u^f_1,u^f_2\}$.

\end{enumerate}

There are three very interesting things to note about this
algorithm. First of all, note that this algorithm actually tests the
performance of the non-linear law, i.e., of one of the best
representations of the model intensity profiles that we have
available, because those coefficients are the ones that are used as
inputs for the algorithm which are then compared to observations. This
is very interesting, as it gives us a simple tool to asses how well
our model atmospheres do at predicting real, observed, intensity
profiles through transit lightcurves. The second thing to note is
that, because the non-linear law is the law being used as input, the
actual method for obtaining the LDCs
\cite[i.e., flux-conservation method, least-squares, etc., see ][for a
discussion on the differences on the obtained coefficients by those
methods for the case of the quadratic law]{howarth2011} is irrelevant
in practice, because all those methods give the same coefficients in
the case of the non-linear law, as shown by \cite{claret2000}. Finally, 
it is very important to note that this algorithm deals mainly with the 
bias introduced by modelling a complex intensity profile 
(e.g., one following a law close to the non-linear) with a simple,
two-parameter law such as the quadratic one in the transit fitting 
procedure. This can be verified by noting that small changes in the input 
non-linear LDCs (e.g., obtained from stars with similar temperatures, 
metallicities or gravities) give rise to larger changes between the 
modelled ($u_i$) and photometrically observed ($u_i^*$) LDCs than changes 
in the transit parameters. This means that although it is true 
that this algorithm deals with the bias associated with transit geometry, 
differences in this mapping between systems are currently mainly dominated 
by stellar parameters rather than by the actual transit geometry of a given system.

Although a big step forward in how to properly compare limb-darkening
coefficients, the SPAM algorithm has a few shortcomings: (1) the
geometric parameters are fitted in step (iii), which is inconvenient
as one would want to fix them, because the objective of the algorithm
is to perform the mapping $(u_1,u_2)\to (u_1^*,u_2^*)$ given a
geometric setting; and (2) the algorithm does not account for
uncertainties on the geometric parameters of the transit lightcurve,
$\vec{\theta}_p$, and/or for uncertainties on the stellar parameters,
$\theta_*$. The latter is a problem because, as stated above, changes in the
geometry of the transit lead to changes in the limb-darkening
coefficients obtained from the SPAM algorithm, while errors on the
stellar parameters can lead to significantly different coefficients
$c_1, c_2, c_3, c_4$ used in step (ii) to generate the synthetic
lightcurves.
%
%
Ideally one does not obtain only best-fit transit parameters for the
geometric parameters, but also their posterior probability
distribution functions (PDFs) given the data via, e.g., a Markov Chain
Monte Carlo (MCMC) algorithm, which is information that we want to
use. In most cases an MCMC approach for obtaining stellar parameters
is not practical and therefore posterior distributions for these
parameters are not usually available. One can still fit an ad-hoc
distribution to a given set of stellar parameters and their respective
errors (e.g., a Gaussian for the case of a parameter with symmetric
error bars), and use that as an approximation to the posterior
distribution of the stellar parameters. We propose here a modified
version of the SPAM algorithm that takes into account uncertainty
information. We term this algorithm Monte-Carlo SPAM algorithm
(MC-SPAM), and it consists of the following three steps:

\begin{enumerate}

\item Fit a transit lightcurve and obtain the posterior PDFs given the
  data for the geometric parameters, i.e., $p(\vec{\theta}_p |
  \textnormal{data})$, with the corresponding posterior for the
  LDCs,
  $p(\vec{u}^f|\textnormal{data})$. Obtain the posterior distribution
  of the stellar parameters (e.g., via spectroscopic observations),
  $p(\vec{\theta}_*|\textnormal{data}_2)$\footnote{Note that we make
    explicit the fact that, in most cases, the dataset used to
    constrain the geometric parameters of the transit and the
    dataset used to constrain the stellar parameters is different. The
    algorithm does not rely on this fact, however.}.
 
\item Draw a set of geometric parameters $\vec{\theta}_p^d$ and
  stellar parameters $\vec{\theta}_*^d$ from the posterior
  distribution $p(\vec{\theta}_p,\vec{\theta}_* |
  \textnormal{data},\textnormal{data}_2)= p(\vec{\theta}_p |
  \textnormal{data}) p(\vec{\theta}_*| \textnormal{data}_2)$. Apply
  the original SPAM algorithm to these parameters and obtain
  $\vec{u}^{*,d}=\{u_1^{*,d},u_2^{*,d}\}$, the SPAM limb-darkening
  coefficients for the given draw.

\item Repeat step (ii) to obtain a Monte-Carlo sample of the
  model/photometric LDCs $\vec{u}^*$, which can
  now be directly compared to the observed LDCs
  $\vec{u}^f$.

\end{enumerate}

Because in most cases no MCMC chains are available for the stellar
parameters, in order to sample directly from
$p(\vec{\theta}_*|\textnormal{data}_2)$, we assume independence among
the stellar parameters and sample the parameters directly from their
respective posterior marginal distributions, i.e.,

\begin{eqnarray*}
p(\vec{\theta}_*|\textnormal{data}_2) = \prod_{i=1}^4 p(\theta_{i,*} |\textnormal{data}_2),
\end{eqnarray*}

\noindent where
$\{\theta_{1,*},\theta_{2,*},\theta_{3,*},\theta_{4,*}\} =
\{T_\textnormal{eff}, \log g, [M/H], v_\textnormal{turb}\}$. For the
modelling of the posterior marginal distributions, we assume the
quoted estimates and errors come from an analysis done on a $\chi^2$
surface, which is essentially an analysis of the likelihood
$\mathcal{L}(\chi^2)$ and, thus, can be approximated by an analysis of
the posterior distribution of the parameters, $p(\vec{\theta}_{*}
|\textnormal{data}_2)\sim \mathcal{L}(\chi^2)$, if one assumes flat
priors for them. In practice, we model each marginal distribution by a
Gaussian in the case of symmetric error bars, where the mean is set to
the best-fit value of the parameter and its variance to the square of
the error. For asymmetrical error bars, we assume a skew-normal
distribution \citep{azzalini85}, which is the natural choice for a
``normal-like'' distribution with lack of symmetry. Details on the
method we use to fit the parameters of this distribution given an
estimate of a parameter and a set of asymmetrical error bars, as well
as how to sample from the resulting distribution are given in
Appendix~C.
We note that although $\log g$, $v_\textnormal{turb}$ and
$T_\textnormal{eff}$ are positive quantities and our treatment can in
principle allow negative values, in practice, the parameter
uncertainties do not allow such values to be sampled and even if they
did, our algorithm is capable of detecting and discarding those values
as invalid.

Our MC-SPAM algorithm is available at
GitHub\footnote{\texttt{http://www.github.com/nespinoza/mc-spam}}. We
now use it to compare our estimates of the observed limb-darkening
coefficients to a set of {\em Kepler} planets.

\subsection{Comparing model to observed LDCs
  using {\em Kepler} data}

Photometrically derived LDCs from fits to {\em
  Kepler} transit lightcurves were obtained from various sources in
the literature in order to retrieve LDCs for
stars with different parameters that host transiting planets with
different geometries. Because the posterior distributions for the
parameters of those systems are not published, we choose to use the
same parametrization used for the stellar parameters in order to
sample values given their published quantiles (i.e., in case of
symmetrical errorbars we assume the posterior, marginalized
distribution of each parameter is a Gaussian, while for the case of
asymmetrical errobars we assume the posterior is best described by a
skew-normal distribution). We note that the uncertainties associated 
with the transit parameters had
negligible influence on the retrieved estimates of the limb-darkening
coefficients obtained with our MC-SPAM algorithm, which were mainly
dominated by the stellar parameter uncertainties.

We took data from the high signal-to-noise, low-impact parameter
sample of \cite{muller2013} where we additionally removed the objects that had
impact parameters larger than $b=0.5$, which were Kepler-43b
($b=0.65$), Kepler-45b ($b=0.6$), Kepler-7b ($b=0.556$), Kepler-8b
($b=0.72$), KIC 5357901b \citep[KOI-188b,][$b=0.6$]{hebrard2014},
Kepler-41b ($b=0.54$) and Kepler-15b ($b=0.56$). In order to be as
conservative as possible, we also removed all the objects which have
not been confirmed as planets to date, including Kepler-71b, for which
no spectroscopic confirmation has been published so far that can rule
out the possibility of it being a low-mass star, as \cite{howell2010}
can only constrain its mass to be less than $0.1$ solar masses. We
also remove Kepler-3b (HAT-P-11b) from our sample because its host
star has been shown to have a significant amount of activity
\citep{fraine2014} and thus the LDCs might be
significantly biased \citep{csizmadia2013}. In addition, we add
Kepler-93b to our sample \citep{ballard2014}, the recently validated
Jupiter-sized planets KOI-206b and KOI-680b \citep{almenara2015} and,
at the expense of larger errors on the estimated limb-darkening
coefficients but in order to expand the effective temperatures sampled
in this work, we also add the newly confirmed planets with low
impact-parameters Kepler-186f, Kepler-296f, Kepler-296e, Kepler-436b,
Kepler-439b, Kepler-440b, Kepler-441b, Kepler-442b and Kepler-443b
\citep{torres2015}. Table~\ref{planet_parameters} summarizes the
best-fit transit parameters for each of these planets, while
Table~\ref{host_star_parameters} summarizes the host-star
parameters. Table~\ref{mc_spam_results_table} presents both the model
LDCs, $(u_1,u_2)$, and the
$\vec{u}^*=(u_1^*,u_2^*)$ coefficients, obtained using $1000$ samples
from our MC-SPAM algorithm for each target. We used both the ATLAS and
PHOENIX models with the methods discussed in Section~2. A
microturbulent velocity of $2$ km/s was assumed for stars for which no
measurement of this parameter was published.

\begin{figure*}
\includegraphics{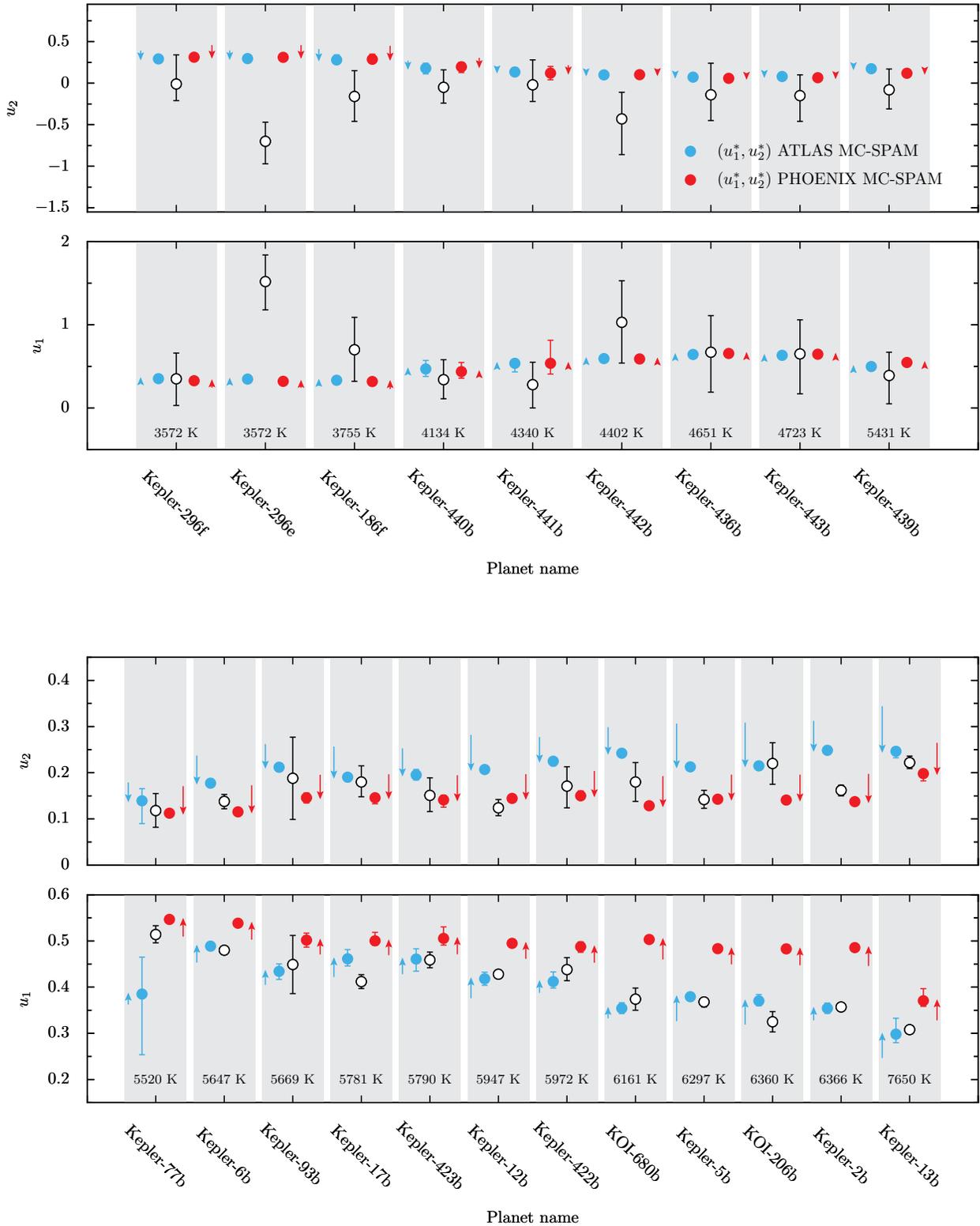}
\caption{High (bottom panels) and low (upper panels) precision
  quadratic LDCs derived from transit
  photometry for several exoplanets (white datapoints) and MC-SPAM
  model LDCs $(u_1^*,u_2^*)$ using ATLAS (blue,
  to the left of each datapoint) and PHOENIX (red, to the right of
  each datapoint) models. The blue and red arrows next to the MC-SPAM
  results represent the mapping $u_i \to u_i^*$, i.e., from the original 
  model LDCs obtained from fits to the intensity 
  profiles (in practice obtained using the non-linear coefficients through 
  equation \ref{eq:limcoeffs}) and our MC-SPAM estimates. The temperature 
  of the host star of each system is indicated above each of the planet names, 
  inside the figures. Note the change in scale between the upper and lower panels.}
\label{mcspam_all_planets}
\end{figure*}

\begin{figure*}
\includegraphics{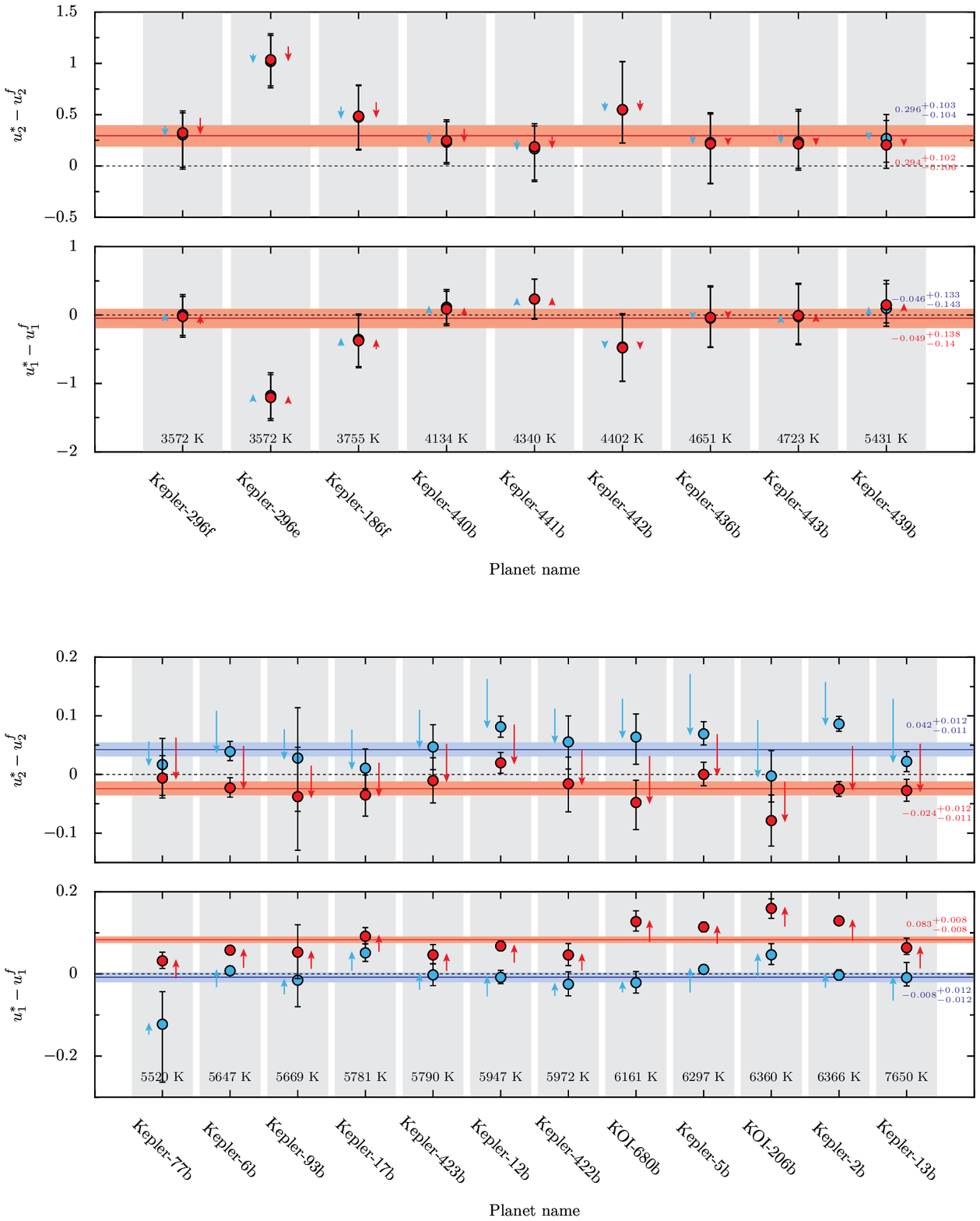}
\caption{Differences between the observed LDCs,
  $(u_1^f, u_2^f)$, shown in Figure~\ref{mcspam_all_planets} as white
  datapoints and the MC-SPAM estimates, $(u_1^*, u_2^*)$, using the
  ATLAS (blue) and PHOENIX (red) models, which are the blue and red
  points next to the white datapoints in
  Figure~\ref{mcspam_all_planets}. The arrows represent the
  effect of the MC-SPAM algorithm in the observed difference between
  $(u_1^f, u_2^f)$ and the original model LDCs
  obtained from fits to the intensity profiles. The blue bands
  indicate the 68\% bands around the mean of the differences
  using the ATLAS results (with the median of this indicated by the
  blue solid line), while the red bands show the same for the PHOENIX
  models (with the red solid line indicating the median of this
  distribution). These medians and the associated 68\% values are also
  indicated next to each band (note that in the upper panels both bands 
  overlap). The dashed black line marks zero, for
  reference.}
\label{mcspam_all_planets_differences}
\end{figure*}

Figure~\ref{mcspam_all_planets} shows the published limb-darkening
coefficients, $(u_1^f,u_2^f)$, as white points with errorbars.  For
easier visualization, the planets have been divided into two groups:
the ones with low precision (upper panels) and high precision (lower
panels) limb-darkening measurements; note that these also separate the
cooler (upper panels, $T_\textnormal{eff} = 3572-5431$ K) and the
hotter ($T_\textnormal{eff} = 5520-7650$ K) stars in the sample. At
the sides of each datapoint, the MC-SPAM results, $(u_1^*,u_2^*)$ for
each system using the ATLAS models (left, blue points) and PHOENIX
models (right, red points) are shown. The arrows show the changes from
the median of the {\em model} quadratic LDCs
sampled by the MC-SPAM algorithm, $(u_1,u_2)$ (i.e., the values
obtained directly from the intensity profiles; in this case, they are
the ``limiting coefficients'' obtained using the non-linear law), to
the median of the resulting MC-SPAM values,
$(u_1^*,u_2^*)$. Figure~\ref{mcspam_all_planets_differences} shows the
differences between the measured LDCs and the
MC-SPAM values, with colored bands indicating the 68\% band of the
mean of those differences for the case of the ATLAS (blue bands) and
PHOENIX (red bands) models. Kepler-296e was omitted from the
calculation of the distribution of the mean of the differences (see
below).

In general, just as we observed in our simulation shown in
Figure~\ref{simulation_howarth}, the shifts between the $(u_1,u_2)$
and $(u_1^*,u_2^*)$ coefficients produced by the MC-SPAM are larger
for the $u_2$ coefficients than for the $u_1$ coefficients, and the
effect is to correct the model underestimation of the $u_1$
coefficients and the model overestimation of $u_2$; in other words,
the effect of the mapping $u_1 \to u_1^*$ is to increase the value of
$u_1$, while the effect of the mapping $u_2 \to u_2^*$ goes in the
opposite direction. As observed in Figures~\ref{mcspam_all_planets}
and~\ref{mcspam_all_planets_differences}, this mapping is in general
very effective at better describing the observations, especially if the
ATLAS model intensity profiles are used as inputs. It is interesting to
note that the errors given by MC-SPAM are always dominated by the
errors on the stellar parameters; this fact is especially evident for
Kepler-77, whose uncertainty in the microturbulent velocity of $\pm
0.3$ km/s leads to larger errorbars on the results using the ATLAS
models. This highlights the importance of estimating this parameter
directly from spectroscopic measurements.

From the sample of low precision LDCs, which is
also the cooler star sample (upper panels of
Figures~\ref{mcspam_all_planets}
and~\ref{mcspam_all_planets_differences}), one can see that, without
taking into account Kepler-296e, the agreement with the data is very
good for ATLAS and PHOENIX model atmospheres for the $u_1$
coefficients, which shows a mean difference of
$-0.046^{+0.133}_{-0.143}$ if one uses the ATLAS models and
$-0.049^{+0.138}_{-0.140}$ if one uses the PHOENIX models, both of
which are consistent with a zero mean difference.  There is a barely 
significant offset of the $u_2$ coefficients ($2.8\sigma$ for the ATLAS
models, $2.9\sigma$ for the PHOENIX models), with the model values
apparently systematically overestimating those coefficients by a mean
value of $\sim 0.3$. For the sample of high precision limb-darkening
coefficients, which is also the hotter star sample (lower panels in
those figures), the agreement of the $u_1$ coefficients is very good
for the ATLAS models, with a mean difference of
$-0.008^{+0.012}_{-0.012}$ which is consistent with zero, but poor for
the PHOENIX models, whose mean difference is
$0.083^{+0.008}_{-0.008}$, inconsistent with zero by more than
$10\sigma$. Note that this bias is very evident and more prominent for
stars hotter than $6000$ K, where the model values overestimate the
LDCs by $\sim 0.1$. For the $u_2$ coefficients,
both model atmospheres show slightly significant biases, with the
PHOENIX models doing a better job at predicting the observed
LDCs with a mean difference of
$-0.024^{+0.012}_{-0.011}$, which is $2\sigma$ away from zero, and
with the ATLAS models showing a mean difference of
$0.042^{+0.012}_{-0.011}$, which is $3.8\sigma$ away from zero.

As a final note, there is one particular object worth discussing in
detail, Kepler-296e, which has LDCs which
deviate more than $2\sigma$ from those of systems with host stars of
similar stellar parameters, including Kepler-296f, which according to
\cite{torres2015} orbits the same host star in an orbit almost two
times farther away from it. As we can see in
Figures~\ref{mcspam_all_planets}
and~\ref{mcspam_all_planets_differences}, the coefficient changes
induced in Kepler-296e due to the geometry of the transit are not
expected to be very different from those of Kepler-296f, so the
geometry of the system cannot explain the differences on the observed
LDCs. Activity could, in principle, produce
significant biases on the LDCs through
unnoculted spots \citep{csizmadia2013}, but it seems unlikely that
activity affected only one of the observed transits. Because Kepler-296 is known to be a
tight binary \citep{lissauer14}, one might be tempted to think that Kepler-296e maybe did 
not orbit the same star as Kepler-296f as claimed by \cite{torres2015}, but its companion. 
However, both of the stars in the system have actually very similar spectral types and, thus, very similar LDCs, 
which implies that the observed LDCs are actually very different compared to \textit{both} 
stars in the system. An analysis of other alternative hypotheses is out of the 
scope of this work, but we note that these
are the kind of analyses that can be performed from measuring,
comparing and interpreting LDCs from transit
lightcurves using our MC-SPAM algorithm.

\section{The effect of using fixed LDCs in transit fitting}

In the past section, we showed that, as first noted by
\cite{howarth2011},  LDCs extracted from
fits to intensity profiles of stellar model atmospheres, $(u_1,u_2)$,
are not directly comparable to the LDCs
obtained from transit photometry, $(u_1^f, u_2^f)$. This is due to the
fact that the two optimization procedures are significantly different
from one another and, thus, a geometry dependent mapping using
synthetic lightcurves has to be carried out in order to obtain the
coefficients $(u_1^*,u_2^*)$ that can then be compared to the observed
LDCs. This implies that even if one could
measure with excellent precision the intensity profile of a given
star, obtain its LDCs with that profile, and
then measure those coefficients from transit photometry, also with
excellent precision, there is an expected bias between the two
sets of coefficients. This in turn means that if one fixes the
LDCs obtained from the intensity profile in the
transit fitting procedure, then one is using potentially biased
coefficients that can then lead to biased transit
parameters\footnote{The level of bias will depend, among other
  factors, on the bandpass. In particular, issues noted in this paper
  should in general be less severe in the infra-red.}. A strategy to
avoid this bias could be to let the LDCs as
free parameters in the fit.  However, we also expect a bias on the
transit parameters in this case if, as it is usually done, the
intensity profile is modelled with a quadratic law which we we have
seen is not able to accurately describe the full intensity profiles. 

In addition to the above mentioned problems, there is the issue
related to our imperfect knowledge of the underlying, ``true'',
intensity profile. As we saw in \S3, this is currently an issue as 
our models are not able to reproduce the observed LDCs with sufficient 
accuracy. On top of this, according to our results in \S 2, there are 
differences even between our own modelling of those profiles both 
between different model atmospheres and between the different methods 
used to derive the LDCs from them.

In order to explore these sources of bias, in this section we perform
simulations to study the possible biases introduced by our
limb-darkening assumptions on the retrieved transit parameters, using
transit lightcurves generated with the formalism of \cite{ma2002}.

\subsection{A simulation study}

In order explore the effect on the retrieved transit parameters of
fixing or having the the LDCs as free
parameters, we simulate transit lightcurves with unit period, circular
orbits and an assumed intensity distribution for the host star of the
transiting planet. The choice of units such that $P=1$ is just for
convenience of sampling directly in phase and has no consequences for
what follows. The geometric parameters of the transit we can retrieve
from our simulated light curves are the planet-to-star radius ratio,
$p=R_p/R_*$, the semi-major axis to stellar radius ratio, $a_R =
a/R_*$, and the inclination of the orbit, $i$. The simulations were
performed as follows. First, based on the data from all transiting 
planets discovered to date, we choose to
generate synthetic transit lightcurves for planets with all the
combinations of parameters $\{a_R, p\}$ with values $a_R = \{3.27,
3.92, 4.87, 6.45, 9.52, 18.18, 200\}$ and $p = \{0.01, 0.06, 0.11,
0.16, 0.21\}$. In order to explore the effect of different impact
parameters, $b=\cos(i)a_R$, we also varied $b$ from 0 to 0.9 in steps
of $0.1$. This defines 350 different orbital configurations for our
simulations. For each orbital configuration, we simulated $100$
noiseless, uniformly sampled lightcurves with $1000$ in-transit points
and $400$ out-of-transit points each, whose initial times were randomly
perturbed.  We first assume perfect knowledge of the underlying
intensity profile by generating the transits using the non-linear law
with the coefficients $\{c_1,c_2,c_3,c_4\}$ obtained in Section~2 for
models with $\log g =4.5$, solar metallicity, $v_\textnormal{turb}=2$
km/s and effective temperatures between $3500$ K and $9000$ K. We use
the ATLAS models, the fitting method of CB11 and the {\em Kepler}
bandpass. Once we generate a simulated light curve, we retrieve its
transit geometric parameters using constrained non-linear least
squares with the Levenberg-Marquardt algorithm using the
\texttt{lmfit}
package\footnote{\texttt{http://cars9.uchicago.edu/software/python/lmfit/}}.
We perform the fit in two ways: (1) fixing the limb-darkening
coefficients to the ones given by the quadratic law for the given star
(using the limiting coefficients defined in Section~2); and (2)
leaving the LDCs as free parameters in the
fit. In the latter case, we fit for the parameters $q_1 = (u_1 +
u_2)^2$ and $q_2 = u_1/2(u_1+u_2)$, where the parameters $q_1$ and
$q_2$ are constrained to be in $(0,1)$, in order to obtain physically
sound solutions in our non-linear least-square fits
\citep{kipping2013}. For each combination of the geometrical
parameters, the median of the fitted parameters obtained from the 100
generated lightcurves, along with the corresponding errors, are
reported.

\subsection{The case of central transits}

\begin{figure*}
\includegraphics{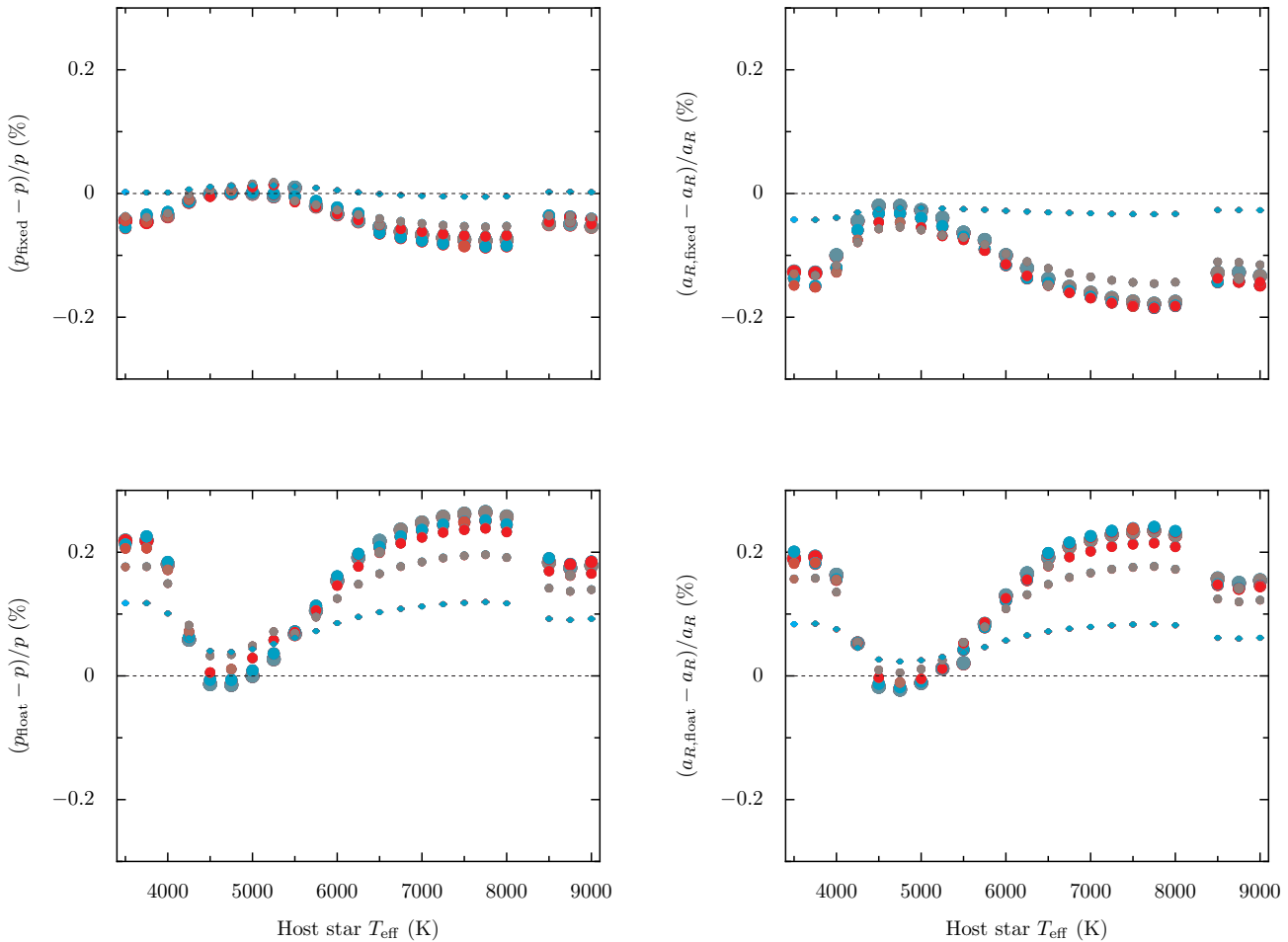}
\caption{Biases in the recovered planet-to-star radius ratio $p=R_p/R_*$ and the semi-major
  axis-to-stellar radius ratio $a_R = a/R_*$ as a function of temperature obtained from the
  simulations described in the text. The upper panels show results when fixing the
  LDCs, while the lower panels show the results when letting them to float in the 
  transit light curve fit. The size of the points denote the
  input value of $p$ ($p=0.01$, small, $p=0.21$ big points), while the
  color of the points represent the input value of $a_R$ ($a_R =
  3.27$, blue points, $a_R=200$ red points).}
\label{simulation_bias}
\end{figure*}

Figure~\ref{simulation_bias} shows the results of our simulations for
the simplest case of a central transit, where the percentual bias
induced on $p$ and $a_R$ is shown as a function of the effective
temperature of the host star used to extract the limb-darkening
coefficients; the upper panel shows the bias induced when fixing the
LDCs to model values and the lower panel shows
the bias induced when letting those be free parameters in the
optimization procedure (the retrieved inclinations in this case were
not reported as they all arrived at the input values). The sizes and
colors of the points represent the input values of $p$ and $a_R$,
respectively, with smaller, bluer points representing the lower values
of both $p$ and $a_R$ ($p=0.01$, $a_R=3.27$), and big, red points
representing the highest input values of those parameters ($p=0.21$,
$a_R=200$). Note that the error bars are plotted, but are smaller than 
the smallest points in the plot. Note also that some points overlap.

We infer from our simulations that the biases are small for central
transits (maximum of $\sim 0.2\%$ on $p$ in the case of fitting for
the coefficients, $\sim -0.05\%$ when fixing them), although
significant for several exoplanets: from a query done to the NASA
Exoplanet Archive\footnote{Query done on 2015/01/25.}, 326 Kepler
Objects of Interest (KOIs) have quoted uncertainties on $p$ lower than
$0.2\%$ and for which the effects of this bias are important. Of those
systems 57 are confirmed exoplanets and 113 show quoted uncertainties
lower than $0.05\%$ (22 of which are confirmed exoplanets). It is
interesting to note that the bias seems to be more important for
deeper transits (bigger points in the Figure), and the effect is to
retrieve deeper transits and larger distances to the host star when
fitting for the coefficients, while the opposite effect is introduced
when fixing the coefficients, estimating slightly shallower transits
and smaller distances to the host star than the real
ones. Furthermore, the bias is clearly larger when fitting for the
coefficients than when fixing them, which suggests that, if the underlying 
limb-darkening model is accurate at a higher level than the error made by fitting the 
coefficients, then the best strategy is fixing the LDCs to their model
values.

We note that the shape that the biases have as a function of effective
temperature  are very similar between the different
strategies employed to fit the transit lightcurves. For the
{\em{Kepler}} bandpass, the difference
between the $u_i$ (LDCs obtained from model intensity profiles) and the
$u_i^*$ (same coefficients recovered from transit fitting) observed in our
experiment in the introduction of \S 3 (Figure \ref{simulation_howarth}) follow
a similar shape with temperature as that observed for the biases, that
is, starting
at the cooler temperatures the offset is larger, decreasing until it reaches a
minimum offset at $T_\textnormal{eff}=4750$ K, then gradually
increasing again for hotter
host stars, slightly decreasing around $T_\textnormal{eff}=8500$ K.
This means that
the actual mapping $u_i\to u_i^*$ is less severe for temperatures
around $T_\textnormal{eff}=4750$ K, which is one of the reasons why we
observe a smaller bias in the retrieved transit parameters around
those temperatures
in our experiments. Additionally, the performance of the quadratic law
fit is also optimal around $T_\textnormal{eff}=4750$ for this
particular bandpass, becoming slightly worse for cooler and hotter
stars, a fact already discussed in \S 2.1.2 (Figure
\ref{lds_kepler_methods}). The two points above serve to explain the
observed shape of the biases with temperature and in particular the
observed minimum of the biases at $T_\textnormal{eff}=4750$ K.

\subsection{The case of low and high impact parameter transits}

\begin{figure*}
\includegraphics{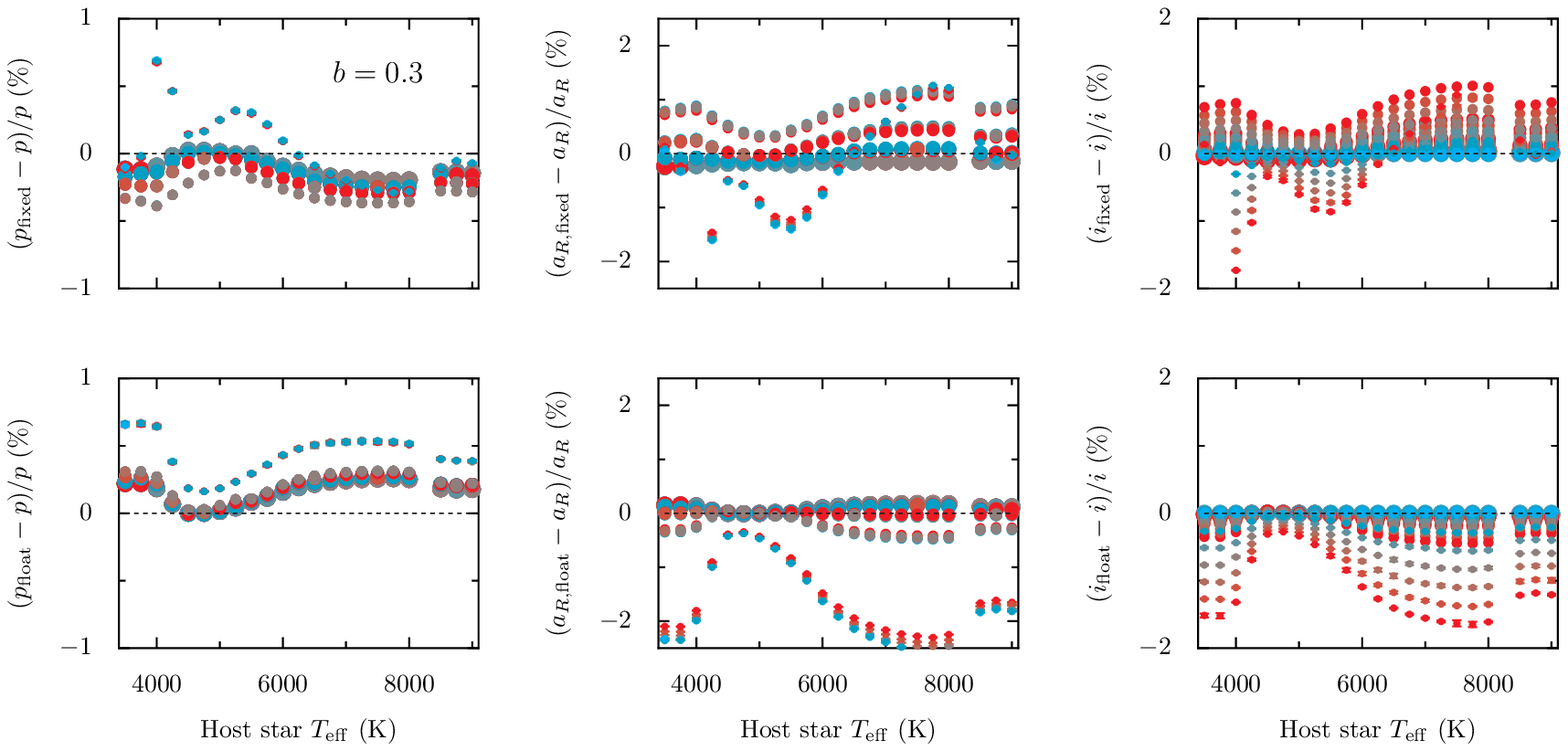}
\caption{Same as Figure~\ref{simulation_bias}, but with an impact
  parameter of $0.3$ and an additional panel showing the bias induced
  on inclination $i$.  Note that the scale is different than that of Figure
  \ref{simulation_bias}.}
\label{simulation_bias_varying_b03}
\end{figure*}

\begin{figure*}
\includegraphics{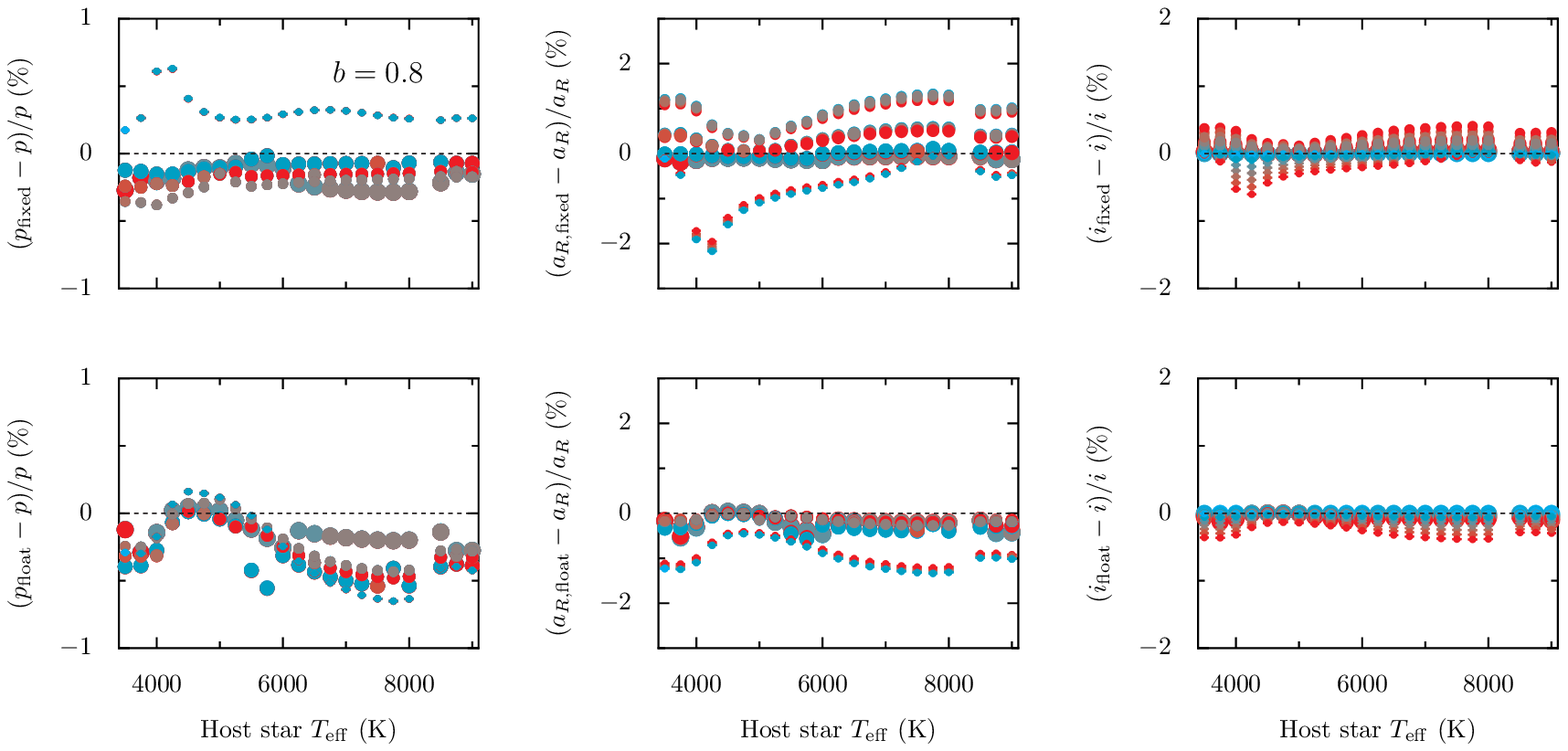}
\caption{Same as Figure~\ref{simulation_bias_varying_b03}, but with an impact parameter of $0.8$. Note that the scale is different than that of Figure~\ref{simulation_bias}, but the same as that of Figure~\ref{simulation_bias_varying_b03}.}
\label{simulation_bias_varying_b08}
\end{figure*}

Figures~\ref{simulation_bias_varying_b03}
and~\ref{simulation_bias_varying_b08} show the results of our
simulations for the cases of low ($b=0.3$) and high ($b=0.8$) impact
parameters. We can see that the value of $b$ strongly affects the
observed bias in the retrieved transit parameters, showing biases as
large as $\sim 1\%$ for $p$, and $\sim 2\%$ for $a_R$ and $i$. These
biases are larger than for the case of central transits and,
therefore, more important.  A query to the NASA Exoplanet Archive
returns 933 KOIs with uncertainties better than $1\%$ on $p$, 164 of
which are confirmed exoplanets. Furthermore, the impact parameter not
only modifies the order of magnitude of the observed bias, but also
modifies the trends observed in the case of central transits. For
example, in this case, the effect is more important for shallow
transits than for deep transits at most temperatures.

One very interesting fact about our simulations is that although for
low impact parameters the bias seems to be larger for $a_R$ and $i$
when fitting for the LDCs, for high impact
parameters this effect is reversed, showing higher biases in those
parameters when fixing the coefficients. The same effect can be seen
for the bias in $p$ in the case of small planets around low
temperature stellar hosts. This is in agreement with the results of
\cite{howarth2011}, who showed that the difference between
LDCs obtained from model atmospheres and the
ones obtained from transit photometry increases as one increases the
impact parameter of the transit. This implies that the fixed
limb-darkening coefficient strategy should worsen as one increases the
impact parameter, which is what we observe in our
simulations. Therefore, for high impact parameter transits one should
fit for the LDCs if one is interested in
decreasing the bias, which contrasts with the suggestion of
\cite{muller2013} of fixing them for high impact parameter
transits based on the observed increased uncertainty in the retrieved
transit parameters when fitting for the coefficients.  In our view,
the best strategy strongly depends on the quality of the photometry
and the observed geometry: if the retrieved uncertainties when fixing
the coefficients are smaller than the biases shown here, then the best
strategy should be to let the coefficients be free parameters.

\subsection{The effect of an unknown stellar intensity profile}

\begin{figure*}
\includegraphics{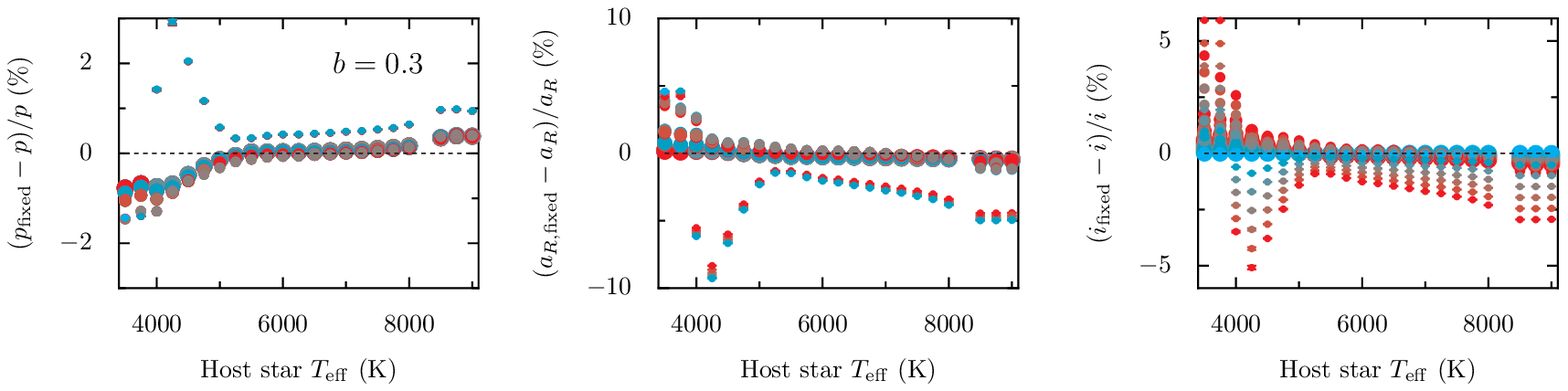}

\caption{Same as Figure~\ref{simulation_bias}, but with an impact
  parameter of $0.3$ and using LDCs obtained from a different 
intensity profile (obtained from the work of CB11) to the underlying one 
(generated using non-linear LDCs from this work). Note that the scale is different 
from that of previous figures.}
\label{simulation_bias_varying_b03_claret}
\end{figure*}

\begin{figure*}
\includegraphics{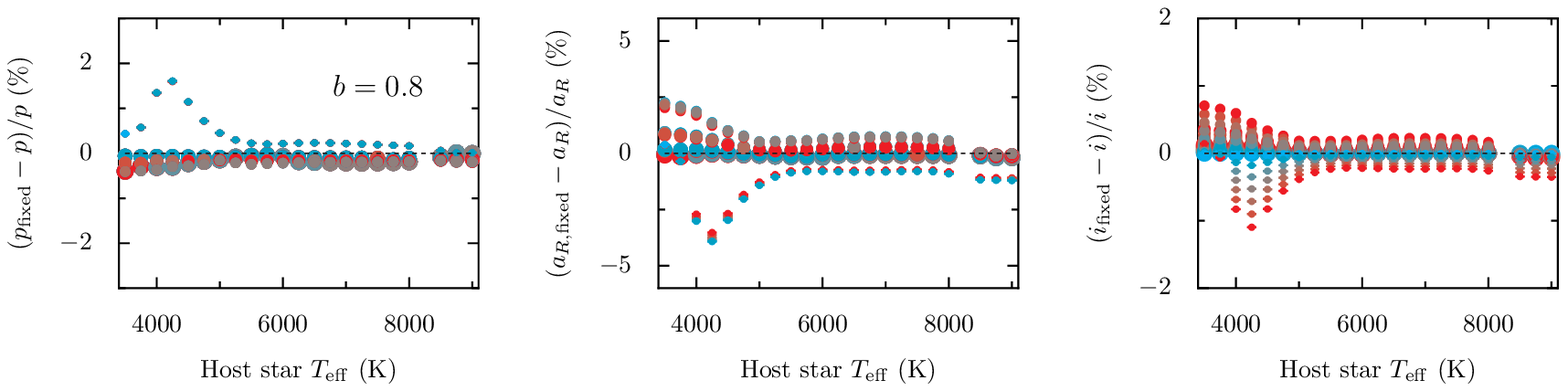}
\caption{Same as Figure~\ref{simulation_bias_varying_b03_claret}, but with an impact parameter of $0.8$. Note that the scale 
between that figure and this one is also different.}
\label{simulation_bias_varying_b08_claret}
\end{figure*}

In order to explore deviations from the known intensity profile,
we make use of previously published limb-darkening coefficients in 
order to simulate systematic offsets from the limb-darkening profile 
being modelled. To this end, we used the same transit lightcurves generated in the past 
sub-sections, generated with our non-linear limb-darkening coefficients, 
but fitted them fixing the limb-darkening coefficients to the values published by CB11. In this way, 
we can explore the 
impact that different methods for obtaining the LDCs  
can have on the retrieved transit parameters which, we note, are 
only lower limits on the actual offsets between the modelled and real 
profiles, as both using our limb-darkening coefficients or the ones published 
by CB11 in our analysis in \S3.2 lead to offsets between observed 
and modelled LDCs of the same order. Figures \ref{simulation_bias_varying_b03_claret} 
and \ref{simulation_bias_varying_b08_claret} show the results of our experiments.

As can be seen from our results, the biases follow a similar shape to the difference 
of the observed quadratic LDCs between our work and that of CB11 in Figure \ref{lds_kepler}; 
that is, as expected, the biases are larger where the calculated LDCs show the largest differences. 
Also, the effect is again more pronounced for smaller values of $p$, with maximum 
biases on the order of $3\%$ for planets around stars with $T_\textnormal{eff}=4000$ K with moderate impact 
parameters ($b=0.3$), temperature and geometry at which we also observe biases on $a_R$ on the order 
of $\sim 10\%$, as well as biases on the order of $5\%$ on $i$. A query to the NASA Exoplanet Archive 
returns 2221 KOIs with uncertainties better than $3\%$ on $p$, 491 of which are confirmed exoplanets. 
This is a clear illustration of the importance of our limb-darkening assumptions.

\section{Discussion}

\subsection{Lessons learned from the comparison of model-to-observed LDCs}
In Section \S 3, we compared observed LDCs to model LDCs using our new 
MC-SPAM algorithm, which not only takes into account the geometry of the 
transit, but also the uncertainties in the transit and host star parameters. 
From these results, it is was apparent that overall the MC-SPAM
results using the ATLAS models do a better job at predicting the
observed $u_1$ LDCs for hotter stars than the
PHOENIX models, with the tables slightly turned in the case of the
$u_2$ coefficients. Both models seem to do a good job at predicting
the $u_1$ coefficients of cool stars, while slightly overestimating
their $u_2$ coefficients. These results suggest that the non-linear 
law in general seems unable to predict the limb-darkening effect with 
the accuracy necessary given the quality of available data and, because 
this is the best representation that we have for the ATLAS and PHOENIX model
atmospheres, this suggests that those models are still not reproducing
the {\em observed} limb-darkening effect with sufficient accuracy. As such, 
these model results have to be taken with care and one must be careful in 
trusting too much the LDCs derived directly from model atmospheres. Our 
view is that one should always be aware of the possible biases introduced by fixing those 
coefficients in practical applications.

\subsection{Limb-darkening and biases in transit parameters: implications for 
exoplanetary science}

The biases uncovered by our experiments in \S 4 are, in general, non-negligible. 
In particular, the large biases shown for the cases 
of non-central transits both when assuming a perfectly known intensity profile 
(\S 4.3) and when assuming deviations from it (\S 4.4) have several implications for
exoplanetary science. First, the biases shown for $p$ have a direct
impact on population studies that rely on accurate measurements of the
planetary radius and their associated uncertainties which, as recently 
shown by \cite{schlaufman2015}, impact the derived conclusions 
of these studies. It also has an impact on present and future
studies that aim to constrain the interior composition of exoplanets
from precise measurements of the planetary radius and mass. For
example, \cite{dorn2015} recently showed that a precision better
than $\sim 2$ \% on the planetary radius is one of the key ingredients
in order to be able to constrain the interior structure of rocky
exoplanets. In addition, the biases for $a_R$ have an impact on
derived quantities such as the calculated incident fluxes on
exoplanets. This parameter, combined with the inclination $i$ of the
orbit, defines in turn the impact parameter, which according to our
results can vary significantly depending on the different
limb-darkening assumptions.

\subsection{Can the biases be corrected and/or avoided?}

As shown in in \S 4, the biases introduced by simple
parametrizations of the limb-darkening effect such as the quadratic
law are significant in comparison to the published parameters
uncertainties for several confirmed and candidate {\em Kepler}
exoplanets. A natural question to ask is: how can one correct for
those biases and how can one avoid them?

The correction of biases in the transit parameters retrieved from a
given fit using the quadratic law would necessarily have to involve
simulations such as the ones shown here around the parameters of
interest. However, an even simpler method would be to fit the original
lightcurve with higher order limb-darkening laws like the
three-parameter or non-linear laws. The problem with this approach 
is that, although it seems obvious to switch to higher order laws once 
the order of magnitude of the bias surpasses the uncertainties in a given parameter, there 
is always an intrinsic bias on the retrieved transit parameters related 
to our potentially imperfect knowledge of the exact shape of the, underlying intensity 
profiles in stars which, as shown in \S 4.4, can lead to large biases if the
coefficients are fixed in the transit fitting procedure. As shown in
Section 3, it appears that better modelling is indeed necessary given
that the observed LDCs are not well reproduced
by the models. This will be an important endeavour to minimize biases
in the parameters estimated from transit light curves, and will be
important to fully exploit data from both present high-precision
missions like {\em Kepler}, and future high-precision missions like
the Transiting Exoplanet Survey Satellite (TESS) or the James Webb
Space Telescope (JWST).

Although a better modelling of stellar atmospheres is an important
task, it is also a long-term one. Given the large amount of data
already obtained by missions such as {\em Kepler} and the advent of
future missions that will focus on retrieving high-precision transit
parameters, it is important to discuss short-term strategies that can
help overcome the biases due to our limb-darkening assumptions shown
in this work. The most natural approach would be to fit the transit lightcurves 
with high order laws letting their parameters to float. However, this will most 
likely be a bad idea for low and medium signal to noise lightcurves due to the 
degeneracy between the parameters fitted in this procedure. One strategy to 
overcome this and the problems stated above could be to select a set of high-precision transit lightcurves with precisely known stellar host
parameters that have enough signal-to-noise as to be able to be fitted
with high order laws such as the three-parameter or non-linear
law. This would in turn allow one to obtain \textit{fitted}
LDCs which could then be used to predict or constrain those coefficients 
for other stars. Although in theory those
fitted LDCs are dependent on the transit geometry, this dependance as 
discussed in Section 3 is rather weak in comparison with the typical 
uncertainties of the host star parameters and, thus, this strategy seems 
to be a promising one for achieving high accuracy and precision measurements 
of transit parameters.

\section{Conclusions}

In this paper, we calculated LDCs using the
{\em Kepler} bandpass for both ATLAS and PHOENIX stellar atmosphere
models, and showed that we cannot reproduce previously published
values. We could not fully resolve the difference due to the fact the
procedures used in the literature are not openly available, but we
believe the differences between the different sources of limb
darkening coefficients available in the literature are due to the use
of different input model atmospheres. We make our codes available for
users to reproduce our results and calculate limb darkening
coefficients for any bandpass in a flexible way.

We showed that different methods used in the fitting stage of the
intensity profiles lead to different LDCs. We
define a set of quadratic law limiting coefficients that are the best
description of the ``real'' underlying intensity distribution in the
sense that they are obtained by sampling uniformly the whole
profile. Among previously published methods using the ATLAS models,
the best is that of \cite{claret11} which by generating additional
points to the ones directly available from the models via spline
interpolation provides a good approximation to a denser sampling of
the underlying profile.
We also point out an important correction that needs to be applied when
dealing with the spherically symmetric PHOENIX models so that the
stellar radius definition is consistent with the plane-parallel ATLAS
models. We provide updated limb darkening coefficients using PHOENIX
models with this correction applied which are now directly comparable
to those derived using ATLAS models.

In order to map model LDCs to those determined
by transit photometry we introduce an algorithm called MC-SPAM
(Monte-Carlo Synthetic-Photometry/Atmosphere-Model), which builds upon
the SPAM algorithm proposed by \cite{howarth2011}. The algorithm takes
into consideration not only the fact that LDCs
obtained from intensity profiles of model stellar atmospheres are not
directly comparable to the coefficients estimated from transit
photometry, but also the fact that all the stellar parameters and
geometrical parameters of the transit have measurement errors.  We use
MC-SPAM to compare limb darkening coefficients predicted by models for
a sample of systems which have had their limb darkening coefficients
determined from {\em Kepler} transit photometry. We show that the
ATLAS and PHOENIX models are not able to fully reproduce the observed
limb-darkening effect for systems with a wide range of stellar
parameters. Finally, we showed that, when using the quadratic law,
fixing and letting the LDCs be free in the 
transit fitting procedures induce biases on the retrieved transit parameters 
that are significant for several candidate and confirmed {\em Kepler} planets, 
even if one assumes a perfectly known stellar intensity profile.

Given our results, we conclude that if one is confronted with a transit lightcurve 
and one decides to use the quadratic limb-darkening law to model the 
limb-darkening effect, the best strategy in order to minimize the bias introduced in the transit 
parameters is to let them float in the transit fitting procedure. Although 
a natural strategy to follow once the biases
introduced by a given limb-darkening law are important for a given
system geometry and precision would be to switch to higher order
laws, one has to be careful if the strategy is to fix the limb-darkening 
coefficients. This is because the strategy would also lead to an underlying 
bias due to our still incomplete knowledge of the
intensity distribution of real stars;
\textit{our understanding of stellar model atmospheres is not good enough to
avoid biases by switching to using fixed model coefficients from
higher order laws}. As a short-term solution to this problem, we
propose to fit a set of high signal-to-noise lightcurves with
precisely known stellar host parameters using a high order law
parametrization such as the three-parameter or non-linear laws in
order to obtain the LDCs directly from data,
which could then be used to predict or at least constrain the values
of those coefficients for other exoplanetary systems.

The results shown in this work imply that the current achievable {\em
  precision} is smaller than the current achievable {\em accuracy} if
one considers only our limb-darkening assumptions. Because of this, we
call the attention to observers to be careful about their
limb-darkening assumptions, and suggest always leaving room for
flexibility on the LDCs in the transit fitting
procedures even when fitting high order laws in order to shield
against biases in the retrieved transit parameters. If precision is
more important than accuracy (e.g., in applications such as
transmission spectroscopy), we suggest performing simulations similar
to the ones shown in this work in order to quantify the order of
magnitude of any biases if the choice is to leave the limb-darkening
coefficients fixed.

\section{Acknowledgments}

N.E. is supported by CONICYT-PCHA/Doctorado Nacional. N.E. \& A.J. acknowledge 
support from the Ministry for the Economy, Development, and Tourism’s Programa Iniciativa
Cient\'ifica Milenio through grant IC 120009, awarded to the
Millennium Institute of Astrophysics (MAS). A.J. acknowledges support from FONDECYT 
project 1130857 and from BASAL CATA PFB-06. We would like to thank the anonymous 
referee for helpful comments, questions and suggestions that helped to improve 
this work. We would also like to thank Pierre Kervella \& Antoine M\'erand who pointed out 
to us the fact that PHOENIX models have a limb definition
inconsistent with that of plane-parallel ATLAS models, Benjamin Rackham for his assistance 
on writing initial versions of the fitting procedures for the PHOENIX models, Antonio
Claret, David Sing \& Tom Evans for useful discussions regarding the
procedures they used for obtaining LDCs, and
Ashley Villar, Jontahan Fraine, Amaury Triaud and Rafael Brahm for useful
discussions regarding the results of this paper.

Some of the data presented in this paper were obtained from the Mikulski Archive for Space Telescopes (MAST). STScI is operated by the Association of Universities for Research in Astronomy, Inc., under NASA contract NAS5-26555. Support for MAST for non-HST data is provided by the NASA Office of Space Science via grant NNX13AC07G and by other grants and contracts. This paper includes data collected by the Kepler mission. Funding for the Kepler mission is provided by the NASA Science Mission directorate.

\clearpage
\begin{landscape} 
 \begin{table}
 \caption{Sample of confirmed Kepler planets for which limb-darkening coefficients have been obtained.}
 \label{planet_parameters}
 \begin{tabular}{@{}lcccccccl}
  \hline\\ 
  \vspace{0.2 cm}
  Planet name & $R_p/R_*$ & $i/^{\circ}$ & $a/R_*$ & $e$ & $\omega$ & $u_1^f$ & $u_2^f$ & Reference \\
  
  \hline\\
  
  \vspace{0.1 cm}
  Kepler-423b & $0.12662^{+0.00029}_{-0.00028}$ & $87.78^{+0.10}_{-0.11}$ & $8.163^{+0.030}_{-0.034}$  & $0$ (fixed) 
  & $0$ (fixed) & $0.459^{+0.017}_{-0.017}$ & $0.151^{+0.038}_{-0.035}$ & \cite{muller2013} \\ 
  
  \vspace{0.1 cm}
  Kepler-77b & $0.09958^{+0.00026}_{-0.00026}$ & $88.00^{+0.10}_{-0.12}$ & $9.749^{+0.054}_{-0.055}$  & $0$ (fixed) 
  & $0$ (fixed) & $0.514^{+0.019}_{-0.018}$ & $0.118^{+0.037}_{-0.036}$ & \cite{muller2013} \\
  
  \vspace{0.1 cm}
  Kepler-17b & $0.13354^{+0.00017}_{-0.00019}$ & $89.71^{+0.29}_{-0.13}$ & $5.707^{+0.010}_{-0.008}$  & $0$ (fixed) 
  & $0$ (fixed) & $0.412^{+0.015}_{-0.015}$ & $0.180^{+0.035}_{-0.032}$ & \cite{muller2013} \\
  
  \vspace{0.1 cm}
  Kepler-6b & $0.09412^{+0.00008}_{-0.00013}$ & $89.39^{+0.26}_{-0.38}$ & $7.551^{+0.029}_{-0.011}$  & $0$ (fixed) 
  & $0$ (fixed) & $0.480^{+0.007}_{-0.007}$ & $0.138^{+0.015}_{-0.016}$ & \cite{muller2013} \\
  
  \vspace{0.1 cm}
  Kepler-422b & $0.09625^{+0.00025}_{-0.00022}$ & $88.09^{+0.05}_{-0.06}$ & $13.826^{+0.073}_{-0.076}$  & $0$ (fixed) 
  & $0$ (fixed) & $0.438^{+0.026}_{-0.024}$ & $0.171^{+0.042}_{-0.047}$ & \cite{muller2013} \\
  
  \vspace{0.1 cm}
  Kepler-12b &  $0.11879^{+0.00013}_{-0.00013}$ & $88.79^{+0.11}_{-0.13}$ & $8.018^{+0.019}_{-0.020}$  & $0$ (fixed) 
  & $0$ (fixed) & $0.428^{+0.007}_{-0.008}$ & $0.124^{+0.018}_{-0.017}$ & \cite{muller2013} \\
  
  \vspace{0.1 cm}
  Kepler-2b & $0.07764^{+0.00004}_{-0.00004}$ & $83.12^{+0.04}_{-0.05}$ & $4.152^{+0.006}_{-0.006}$  & $0$ (fixed) 
  & $0$ (fixed) & $0.357^{+0.007}_{-0.007}$ & $0.162^{+0.011}_{-0.012}$ & \cite{muller2013} \\
  
  \vspace{0.1 cm}
  Kepler-5b &  $0.07972^{+0.00006}_{-0.00009}$ & $89.39^{+0.61}_{-0.20}$ & $6.459^{+0.023}_{-0.008}$  & $0$ (fixed) 
  & $0$ (fixed) & $0.368^{+0.009}_{-0.011}$ & $0.142^{+0.020}_{-0.019}$ & \cite{muller2013} \\
  
  \vspace{0.1 cm}
  Kepler-13b & $0.08553^{+0.00007}_{-0.00007}$ & $85.82^{+0.10}_{-0.12}$ & $4.434^{+0.011}_{-0.010}$  & $0$ (fixed) 
  & $0$ (fixed) & $0.308^{+0.007}_{-0.007}$ & $0.222^{+0.014}_{-0.013}$ & \cite{muller2013} \\
  
  \vspace{0.1 cm}
  Kepler-93b &  $0.014751^{+0.000059}_{-0.000059}$ & $89.183^{+0.044}_{-0.044}$ & $12.496^{+0.015}_{-0.015}$  & $0$ (fixed) 
  & $0$ (fixed) & $0.449^{+0.063}_{-0.063}$ & $0.188^{+0.089}_{-0.089}$ & \cite{ballard2014} \\
  
  \vspace{0.1 cm}
  Kepler-186f &  $0.0205^{+0.0012}_{-0.0013}$ & $89.96^{+0.04}_{-0.10}$ & $178^{+65}_{-21}$  & $0$ (fixed) 
  & $0$ (fixed) & $0.70^{+0.39}_{-0.38}$ & $-0.16^{+0.31}_{-0.30}$ & \cite{torres2015} \\
  
  \vspace{0.1 cm}
  Kepler-296f &  $0.0362^{+0.0022}_{-0.0018}$ & $89.95^{+0.05}_{-0.12}$ & $137^{+34}_{-15}$  & $0$ (fixed) 
  & $0$ (fixed) & $0.35^{+0.31}_{-0.32}$ & $-0.01^{+0.35}_{-0.20}$ & \cite{torres2015} \\
  
  \vspace{0.1 cm}
  Kepler-296e & $0.0297^{+0.0029}_{-0.0037}$ & $89.89^{+0.11}_{-0.26}$ & $71^{+29}_{-10}$  & $0$ (fixed) 
  & $0$ (fixed) & $1.52^{+0.32}_{-0.34}$ & $-0.70^{+0.23}_{-0.27}$ & \cite{torres2015} \\
  
  \vspace{0.1 cm}
  Kepler-436b &  $0.0354^{+0.0024}_{-0.0035}$ & $89.93^{+0.07}_{-0.18}$ & $104^{+34}_{-16}$  & $0$ (fixed) 
  & $0$ (fixed) & $0.67^{+0.44}_{-0.48}$ & $-0.14^{+0.38}_{-0.31}$ & \cite{torres2015} \\
  
  \vspace{0.1 cm}
  Kepler-439b &  $0.02392^{+0.00099}_{-0.00111}$ & $89.95^{+0.05}_{-0.12}$ & $142^{+18}_{-15}$  & $0$ (fixed) 
  & $0$ (fixed) & $0.39^{+0.28}_{-0.34}$ & $-0.08^{+0.25}_{-0.23}$ & \cite{torres2015} \\
  
  \vspace{0.1 cm}
  Kepler-440b & $0.03038^{+0.00112}_{-0.0027}$ & $89.93^{+0.07}_{-0.18}$ & $99.0^{+10.2}_{-9.4}$  & $0$ (fixed) 
  & $0$ (fixed) & $0.34^{+0.24}_{-0.23}$ & $-0.05^{+0.21}_{-0.19}$ & \cite{torres2015} \\
  
  \vspace{0.1 cm}
  Kepler-441b &  $0.0280^{+0.0017}_{-0.0014}$ & $89.97^{+0.03}_{-0.07}$ & $260^{+83}_{-31}$  & $0$ (fixed) 
  & $0$ (fixed) & $0.28^{+0.27}_{-0.28}$ & $-0.02^{+0.30}_{-0.20}$ & \cite{torres2015} \\
  
  \vspace{0.1 cm}
  Kepler-442b & $0.0211^{+0.0019}_{-0.0016}$ & $89.94^{+0.06}_{-0.12}$ & $146^{+80}_{-22}$  & $0$ (fixed) 
  & $0$ (fixed) & $1.03^{+0.50}_{-0.49}$ & $-0.43^{+0.32}_{-0.43}$ & \cite{torres2015} \\
  
  \vspace{0.1 cm}
  Kepler-443b & $0.0304^{+0.0022}_{-0.0022}$ & $89.94^{+0.06}_{-0.13}$ & $151^{+39}_{-22}$  & $0$ (fixed) 
  & $0$ (fixed) & $0.65^{+0.41}_{-0.48}$ & $-0.15^{+0.25}_{-0.31}$ & \cite{torres2015} \\
  
  \vspace{0.1 cm}
  KOI-206b &  $0.06590^{+0.00015}_{-0.00015}$ & $89.21^{+0.52}_{-0.90}$ & $6.44^{+0.62}_{-0.62}$  & $0.119^{+0.079}_{-0.079}$ 
  & $68^{+67}_{-36}$ & $0.325^{+0.022}_{-0.022}$ & $0.220^{+0.045}_{-0.045}$ & \cite{almenara2015} \\
  
  \vspace{0.1 cm}
  KOI-680b & $0.06384^{+0.00020}_{-0.00020}$ & $85.51^{+0.52}_{-0.52}$ & $6.35^{+0.51}_{-0.51}$  & $0.114^{+0.077}_{-0.077}$ 
  & $104^{+36}_{-36}$ & $0.374^{+0.024}_{-0.024}$ & $0.180^{+0.042}_{-0.042}$ & \cite{almenara2015} \\
  \hline
 \end{tabular}
\end{table}

\clearpage
\end{landscape}

\clearpage
\begin{landscape} 
\begin{table}
 \caption{Parameters of the host stars of each of the planets listed in Table \ref{planet_parameters}.}
 \label{host_star_parameters}
 \begin{tabular}{@{}lccccl}
  \hline\\ 
  \vspace{0.2 cm}
  Star name & $T_\textnormal{eff}$/K & $\log g$  (cgs) & $[M/H]^{a}$ & $v_\textnormal{turb}$/km/s & Reference \\
  
  \hline\\
  
  \vspace{0.1 cm}
  Kepler-423 & $5790^{+116}_{-116}$& $4.57^{+0.12}_{-0.12}$ & $0.26^{+0.12}_{-0.12}$ & ... & \cite{endl2014}\\
  
  \vspace{0.1 cm}
  Kepler-77 & $5520^{+60}_{-60}$& $4.40^{+0.10}_{-0.10}$ & $0.20^{+0.05}_{-0.05}$ & $1.8\pm 0.3$ & \cite{gandolfi2013}\\
  
  \vspace{0.1 cm}
  Kepler-17 & $5781^{+85}_{-85}$& $4.53^{+0.12}_{-0.12}$ & $0.26^{+0.10}_{-0.10}$ & ... & \cite{bonomo2012}\\
  
  \vspace{0.1 cm}
  Kepler-6 & $5647^{+44}_{-44}$& $4.236^{+0.011}_{-0.011}$ & $0.34^{+0.04}_{-0.04}$ & ... & \cite{dunham2010}\\
  
  \vspace{0.1 cm}
  Kepler-422 & $5972^{+84}_{-84}$ & $4.50^{+0.10}_{-0.10}$ & $0.23^{+0.09}_{-0.09}$ & ... & \cite{endl2014}\\
  
  \vspace{0.1 cm}
  Kepler-12 & $5947^{+100}_{-100}$ & $4.175^{+0.015}_{-0.011}$ & $0.07^{+0.04}_{-0.04}$ & ... & \cite{fortney2011}\\
  
  \vspace{0.1 cm}
  Kepler-2 & $6366^{+78}_{-80}$ & $4.01^{+0.01}_{-0.01}$ & $0.28^{+0.11}_{-0.11}$ & ... & \cite{lund2014}\\
  
  \vspace{0.1 cm}
  Kepler-5 & $6297^{+60}_{-60}$ & $3.96^{+0.10}_{-0.10}$ & $0.04^{+0.06}_{-0.06}$ & ... & \cite{koch2010}\\
  
  \vspace{0.1 cm}
  Kepler-13A & $7650^{+250}_{-250}$ & $4.2^{+0.50}_{-0.50}$ & $0.2^{+0.20}_{-0.20}$ & ... & \cite{shporer2014}\\
  
  \vspace{0.1 cm}
  Kepler-93 & $5669^{+75}_{-75}$ & $4.470^{+0.004}_{-0.004}$ & $-0.18^{+0.10}_{-0.10}$ & ... & \cite{ballard2014}\\
  
  \vspace{0.1 cm}
  Kepler-186 & $3755^{+90}_{-90}$ & $4.736^{+0.020}_{-0.019}$ & $-0.26^{+0.12}_{-0.12}$ & ... & \cite{torres2015}\\
  
  \vspace{0.1 cm}
  Kepler-296 & $3572^{+80}_{-80}$ & $4.833^{+0.025}_{-0.041}$ & $-0.12^{+0.12}_{-0.12}$ & ... & \cite{torres2015}\\
  
  \vspace{0.1 cm}
  Kepler-436 & $4651^{+100}_{-100}$ & $4.619^{+0.015}_{-0.028}$ & $0.01^{+0.10}_{-0.10}$ & ... & \cite{torres2015}\\
  
  \vspace{0.1 cm}
  Kepler-439 & $5431^{+100}_{-100}$ & $4.514^{+0.035}_{-0.073}$ & $0.02^{+0.10}_{-0.10}$ & ... & \cite{torres2015} \\
  
  \vspace{0.1 cm}
  Kepler-440 & $4134^{+154}_{-154}$ & $4.706^{+0.049}_{-0.016}$ & $-0.30^{+0.15}_{-0.15}$ & ... & \cite{torres2015}\\
  
  \vspace{0.1 cm}
  Kepler-441 & $4340^{+177}_{-177}$ & $4.715^{+0.047}_{-0.024}$ & $-0.57^{+0.18}_{-0.18}$ & ... & \cite{torres2015} \\
  
  \vspace{0.1 cm}
  Kepler-442 & $4402^{+100}_{-100}$ & $4.673^{+0.018}_{-0.021}$ & $-0.37^{+0.10}_{-0.10}$ & ... & \cite{torres2015} \\
  
  \vspace{0.1 cm}
  Kepler-443 & $4723^{+100}_{-100}$ & $4.614^{+0.016}_{-0.029}$ & $-0.01^{+0.10}_{-0.10}$ & ... & \cite{torres2015} \\
  
  \vspace{0.1 cm}
  KOI-206 & $6360^{+140}_{-140}$ & $3.892^{+0.056}_{-0.056}$ & $-0.01^{+0.20}_{-0.20}$ & ... & \cite{almenara2015} \\
  
  \vspace{0.1 cm}
  KOI-680 & $6161^{+94}_{-94}$ & $3.613^{+0.047}_{-0.070}$ & $-0.18^{+0.11}_{-0.11}$ & ... & \cite{almenara2015}\\
  \hline
 \end{tabular}
 
 $^a$ It is assumed that $[M/H]=[Fe/H]$.
\end{table}
\clearpage
\end{landscape}

\clearpage
\begin{landscape} 
\begin{table}
 \caption{Results of the limb-darkening coefficients obtained using the MC-SPAM algorithm.}
 \label{mc_spam_results_table}
 \begin{tabular}{@{}lcccclcccc}
  \hline\\ 
  \vspace{0.2 cm}
  Planet name & $u_1$ (ATLAS) & $u_2$ (ATLAS) & $u_1^*$ (ATLAS) & $u_2^*$ (ATLAS) & $u_1$ (PHOENIX) & $u_2$ (PHOENIX) & $u_1^*$ (PHOENIX) & $u_2^*$ (PHOENIX)\\
  
  \hline\\
  
  \vspace{0.1 cm}
Kepler-423b & $0.419^{+0.023}_{-0.021}$ & $0.263^{+0.013}_{-0.014}$ & $0.457^{+0.022}_{-0.020}$ & $0.197^{+0.009}_{-0.010}$ & $0.463^{+0.021}_{-0.011}$ & $0.206^{+0.007}_{-0.014}$ & $0.504^{+0.021}_{-0.013}$ & $0.142^{+0.009}_{-0.014}$ \\

\vspace{0.1cm}
Kepler-77b & $0.367^{+0.073}_{-0.128}$ & $0.182^{+0.039}_{-0.063}$ & $0.394^{+0.077}_{-0.138}$ & $0.139^{+0.028}_{-0.047}$ & $0.503^{+0.006}_{-0.005}$ & $0.182^{+0.002}_{-0.003}$ & $0.546^{+0.006}_{-0.006}$ & $0.113^{+0.003}_{-0.003}$ \\

\vspace{0.1cm}
Kepler-17b & $0.420^{+0.018}_{-0.016}$ & $0.262^{+0.010}_{-0.011}$ & $0.463^{+0.016}_{-0.016}$ & $0.190^{+0.006}_{-0.008}$ & $0.464^{+0.017}_{-0.009}$ & $0.206^{+0.006}_{-0.012}$ & $0.502^{+0.017}_{-0.010}$ & $0.145^{+0.007}_{-0.012}$ \\

\vspace{0.1cm}
Kepler-6b & $0.448^{+0.008}_{-0.009}$ & $0.245^{+0.006}_{-0.005}$ & $0.488^{+0.008}_{-0.009}$ & $0.178^{+0.005}_{-0.004}$ & $0.495^{+0.006}_{-0.006}$ & $0.185^{+0.004}_{-0.003}$ & $0.538^{+0.006}_{-0.007}$ & $0.116^{+0.005}_{-0.004}$ \\

\vspace{0.1cm}
Kepler-422b & $0.384^{+0.016}_{-0.014}$ & $0.284^{+0.008}_{-0.010}$ & $0.414^{+0.016}_{-0.014}$ & $0.225^{+0.006}_{-0.008}$ & $0.446^{+0.009}_{-0.010}$ & $0.215^{+0.005}_{-0.005}$ & $0.487^{+0.011}_{-0.012}$ & $0.151^{+0.010}_{-0.007}$ \\

\vspace{0.1cm}
Kepler-12b & $0.373^{+0.016}_{-0.016}$ & $0.287^{+0.009}_{-0.009}$ & $0.419^{+0.014}_{-0.013}$ & $0.206^{+0.004}_{-0.005}$ & $0.455^{+0.007}_{-0.008}$ & $0.209^{+0.003}_{-0.002}$ & $0.496^{+0.007}_{-0.008}$ & $0.144^{+0.003}_{-0.003}$ \\

\vspace{0.1cm}
Kepler-2b & $0.324^{+0.011}_{-0.010}$ & $0.319^{+0.005}_{-0.005}$ & $0.354^{+0.012}_{-0.010}$ & $0.248^{+0.003}_{-0.003}$ & $0.438^{+0.006}_{-0.005}$ & $0.210^{+0.002}_{-0.002}$ & $0.486^{+0.006}_{-0.005}$ & $0.137^{+0.002}_{-0.003}$ \\

\vspace{0.1cm}
Kepler-5b & $0.322^{+0.006}_{-0.006}$ & $0.313^{+0.003}_{-0.003}$ & $0.379^{+0.005}_{-0.005}$ & $0.213^{+0.001}_{-0.001}$ & $0.441^{+0.005}_{-0.005}$ & $0.209^{+0.003}_{-0.001}$ & $0.483^{+0.005}_{-0.006}$ & $0.143^{+0.004}_{-0.002}$ \\

\vspace{0.1cm}
Kepler-13b & $0.243^{+0.045}_{-0.021}$ & $0.352^{+0.013}_{-0.028}$ & $0.299^{+0.038}_{-0.019}$ & $0.246^{+0.009}_{-0.012}$ & $0.321^{+0.019}_{-0.012}$ & $0.275^{+0.010}_{-0.013}$ & $0.371^{+0.023}_{-0.013}$ & $0.198^{+0.008}_{-0.015}$ \\

\vspace{0.1cm}
Kepler-93b  & $0.402^{+0.016}_{-0.017}$ & $0.268^{+0.010}_{-0.010}$ & $0.434^{+0.016}_{-0.016}$ & $0.212^{+0.008}_{-0.008}$ & $0.465^{+0.012}_{-0.015}$ & $0.205^{+0.009}_{-0.007}$ & $0.503^{+0.013}_{-0.016}$ & $0.146^{+0.011}_{-0.009}$ \\

\vspace{0.1cm}
Kepler-186f  & $0.259^{+0.042}_{-0.037}$ & $0.420^{+0.025}_{-0.033}$ & $0.318^{+0.039}_{-0.036}$ & $0.303^{+0.031}_{-0.028}$ & $0.219^{+0.035}_{-0.035}$ & $0.466^{+0.024}_{-0.033}$ & $0.302^{+0.033}_{-0.037}$ & $0.309^{+0.031}_{-0.023}$ \\

\vspace{0.1cm}
Kepler-296f  & $0.289^{+0.037}_{-0.032}$ & $0.404^{+0.025}_{-0.028}$ & $0.346^{+0.033}_{-0.032}$ & $0.293^{+0.026}_{-0.018}$ & $0.232^{+0.028}_{-0.027}$ & $0.475^{+0.019}_{-0.020}$ & $0.322^{+0.027}_{-0.031}$ & $0.308^{+0.027}_{-0.014}$ \\

\vspace{0.1cm}
Kepler-296e  & $0.293^{+0.036}_{-0.034}$ & $0.401^{+0.027}_{-0.029}$ & $0.345^{+0.032}_{-0.036}$ & $0.299^{+0.031}_{-0.020}$ & $0.234^{+0.029}_{-0.029}$ & $0.475^{+0.019}_{-0.022}$ & $0.319^{+0.027}_{-0.034}$ & $0.316^{+0.030}_{-0.016}$ \\

\vspace{0.1cm}
Kepler-436b  & $0.617^{+0.013}_{-0.016}$ & $0.115^{+0.011}_{-0.010}$ & $0.644^{+0.015}_{-0.017}$ & $0.071^{+0.014}_{-0.012}$ & $0.619^{+0.011}_{-0.013}$ & $0.117^{+0.009}_{-0.007}$ & $0.657^{+0.012}_{-0.015}$ & $0.057^{+0.012}_{-0.009}$ \\

\vspace{0.1cm}
Kepler-439b  & $0.465^{+0.022}_{-0.021}$ & $0.229^{+0.014}_{-0.015}$ & $0.493^{+0.022}_{-0.021}$ & $0.179^{+0.014}_{-0.014}$ & $0.505^{+0.012}_{-0.012}$ & $0.184^{+0.006}_{-0.005}$ & $0.545^{+0.012}_{-0.013}$ & $0.119^{+0.007}_{-0.007}$ \\

\vspace{0.1cm}
Kepler-440b  & $0.419^{+0.107}_{-0.112}$ & $0.277^{+0.092}_{-0.088}$ & $0.472^{+0.095}_{-0.099}$ & $0.177^{+0.068}_{-0.062}$ & $0.378^{+0.116}_{-0.107}$ & $0.312^{+0.083}_{-0.091}$ & $0.441^{+0.105}_{-0.092}$ & $0.196^{+0.059}_{-0.063}$ \\

\vspace{0.1cm}
Kepler-441b  & $0.479^{+0.076}_{-0.120}$ & $0.219^{+0.090}_{-0.058}$ & $0.520^{+0.062}_{-0.110}$ & $0.146^{+0.062}_{-0.036}$ & $0.456^{+0.117}_{-0.129}$ & $0.260^{+0.088}_{-0.075}$ & $0.507^{+0.105}_{-0.124}$ & $0.166^{+0.070}_{-0.051}$ \\

\vspace{0.1cm}
Kepler-442b  & $0.559^{+0.032}_{-0.047}$ & $0.159^{+0.036}_{-0.025}$ & $0.590^{+0.029}_{-0.038}$ & $0.104^{+0.022}_{-0.017}$ & $0.546^{+0.040}_{-0.061}$ & $0.177^{+0.046}_{-0.034}$ & $0.585^{+0.037}_{-0.052}$ & $0.108^{+0.033}_{-0.023}$ \\

\vspace{0.1cm}
Kepler-443b  & $0.607^{+0.014}_{-0.018}$ & $0.122^{+0.014}_{-0.010}$ & $0.633^{+0.016}_{-0.019}$ & $0.080^{+0.015}_{-0.013}$ & $0.610^{+0.013}_{-0.014}$ & $0.123^{+0.010}_{-0.008}$ & $0.647^{+0.015}_{-0.017}$ & $0.064^{+0.014}_{-0.011}$ \\

\vspace{0.1cm}
KOI-206b  & $0.315^{+0.015}_{-0.013}$ & $0.315^{+0.007}_{-0.006}$ & $0.370^{+0.014}_{-0.011}$ & $0.215^{+0.005}_{-0.003}$ & $0.440^{+0.007}_{-0.008}$ & $0.209^{+0.004}_{-0.004}$ & $0.483^{+0.008}_{-0.008}$ & $0.140^{+0.004}_{-0.003}$ \\

\vspace{0.1cm}
KOI-680b  & $0.328^{+0.010}_{-0.009}$ & $0.305^{+0.004}_{-0.004}$ & $0.353^{+0.012}_{-0.010}$ & $0.242^{+0.004}_{-0.006}$ & $0.451^{+0.005}_{-0.003}$ & $0.207^{+0.001}_{-0.002}$ & $0.503^{+0.005}_{-0.006}$ & $0.128^{+0.006}_{-0.004}$ \\

\hline
\end{tabular}
\end{table}
\clearpage
\end{landscape}

\appendix
\section{Least-squares fits to limb-darkening laws}

Fits of stellar model atmosphere intensities with the laws mentioned in
the introduction require a least-squares procedure to be followed
which, in general, minimizes the quantity

\begin{eqnarray}
\label{chi2}
\chi^2 = \sum_{i=1}^{N} w_i\left(y_i-f(x_i)\right)^2,
\end{eqnarray}

\noindent where $y_i$ are the datapoints to be fitted,
$f(x_i)$ the model for those datapoints, $w_i$ the weight given to
each squared residual $(y_i-f(x_i))^2$ and $N$ the number of
datapoints to be fitted. In our case  we set $w_i=1$ and
the objective is to minimize eq. (\ref{chi2}) using the various
models presented in the introduction of this work.

If no constraint is provided, then the problem is easily solved for all the laws presented because the coefficients are 
linear in the parameters. Our objective then, is to minimize eq. (\ref{chi2}) with $y_i=I(\mu_i)/I(1)$, 
and $x_i=\mu_i$ (the angles at which each of those integrations are done), for the different models $f(\mu_i)$ 
presented in the introduction, obtaining the optimal coefficients in a
least-squares sense. The models have  the form
\begin{eqnarray*}
f(\mu_i) = 1 - \sum_n \theta_n g_n(\mu_i),
\end{eqnarray*}
where $\theta_n$ are the parameters of the laws (e.g., $\theta_n=c_n$ for the non-linear law) and $g_n(\mu_i)$ are the 
functions which make up each law, e.g., in the case of the non-linear law (which contains the linear, square-root and 
three-parameter laws depending on which coefficient $\theta_n$ one sets to zero), $g_n(\mu_i) =
(1-\mu_i^{n/2})$. In order to minimize eq. (\ref{chi2}), we calculate the partial derivatives of $\chi^2$ with respect to the 
different coefficients $\theta_n$ and set them to zero. The calculation is easily found to give
\begin{eqnarray*}
 \frac{\partial \chi^2}{\partial \theta_k} = \sum_{i=1}^N 2(f(\mu_i)-I(\mu_i)/I(1))\frac{\partial f(\mu_i)}{\partial \theta_k} = 0,
\end{eqnarray*}
with
\begin{eqnarray*}
\frac{\partial f(\mu_i)}{\partial \theta_k} = -g_k(\mu_i),
\end{eqnarray*}
which, after rearranging terms, gives the system of $k$ linear equations for the $n$ coefficients $\theta_n$
\begin{eqnarray*}
\sum_{n}\theta_n \alpha_{n,k} = \beta_k,\ k=1,2,...,n,
\end{eqnarray*}
with,
\begin{eqnarray*}
\alpha_{n,k} &=& \sum_{i=1}^N g_n(\mu_i)g_k(\mu_i),\\
\beta_k &=& \sum_{i=1}^N g_k(\mu_i)(1-I(\mu_i)/I(1)),
\end{eqnarray*}
which are trivial to solve. Note also that if we write the linear system as $\mathbf{A}\vec{\theta}=\vec{b}$, with 
$\mathbf{A}_{n,k}=\alpha_{n,k}$, $\vec{\theta}=\{\theta_1,\theta_2,...,\theta_n\}^T$ and $\vec{b}=\{\beta_1,\beta_2,...,\beta_n\}^T$, 
in this case the matrix $\mathbf{A}$ is symmetric, so the system is not only linear but very fast to compute. As an example of how to use 
the above result, for the linear law the only function is $g_1(\mu_i)=1-\mu_i$ so, in this case, there is only one equation for the parameter 
$\theta_1=a$ with parameters
\begin{eqnarray*}
\alpha_{1,1} &=& \sum_{i=1}^N (1-\mu_i)^2,\\
\beta_1 &=& \sum_{i=1}^N (1-\mu_i)(1-I(\mu_i)/I(1)),
\end{eqnarray*}
which gives,
\begin{eqnarray}
\label{linearcoeff}
 \theta_1 = a = \frac{\beta_1}{\alpha_{1,1}} = \frac{\sum_{i=1}^N (1-\mu_i)(1-I(\mu_i)/I(1))}{\sum_{i=1}^N (1-\mu_i)^2}.
\end{eqnarray}
\section{Limiting cases for known target limb-darkening laws}
If we try to fit an arbitrary law when knowing that we are sampling
from a law of the form $I(\mu_i)/I(1)$ (e.g., the non-linear law),
then ``limiting coefficients'' corresponding to $N\to
\infty$ can be obtained. To perform this calculation, we sample
$N$ uniform $\mu_i$ points by defining $\mu_i = (i-1)/(N-1)$, with
$i=1,2,...,N$ (note that this samples $\mu_i$ angles from $\mu_i = 1$
for $i=N$ to $\mu_i=0$ for $i=1$). We introduce this sampling
sums that give the parameters (e.g., eq. (\ref{linearcoeff}) for the
parameter of the linear law) and then take the limit as $N\to
\infty$.

\subsection{Limiting coefficient $a$ for the linear law when sampling from the non-linear law}
We sample $N$ intensity ($I(\mu_i)/I(1)$) and angle ($\mu_i$) pairs from the non-linear law, i.e.,
\begin{eqnarray*}
I(\mu_i)/I(1) = 1-\sum_{n=1}^{4}c_n(1-\mu_i^{n/2}), 
\end{eqnarray*}
with known coefficients $c_n$. We now fit the profile with a linear
law and determine the limiting coefficient $a$ that follows as $N \to
\infty$. We first find the numerator and the 
denominator in equation~(\ref{linearcoeff}). The denominator is easily found to be
\begin{eqnarray*}
\sum_{i=1}^N (1-\mu_i)^2 = \frac{2N^2 - N}{6(N - 1)} = h(N),
\end{eqnarray*}
while the numerator takes the form
\begin{eqnarray*}
\sum_{i=1}^N (1-\mu_i)(1-I(\mu_i)/I(1)) = \sum_{n=1}^{4}c_nS_n,
\end{eqnarray*}
with $S_n = \sum_{i=1}^N (1-\mu_i - \mu_i^{n/2} + \mu_i^{(n+2)/2})$. It is straightforward to show that
\begin{eqnarray*}
 S_1 &=& \frac{N}{2}+\sum_{i=1}^{N}\mu_i^{3/2}-\sum_{i=1}^{N}\mu_i^{1/2},\\
 S_2 &=& \frac{2N^2-N}{6(N-1)}=h(N),\\
 S_3 &=& \frac{N}{2}+\sum_{i=1}^{N}\mu_i^{5/2}-\sum_{i=1}^{N}\mu_i^{3/2},\\
 S_4 &=& \frac{5N^2-4N}{12(N-1)},\\
\end{eqnarray*}
where  we have expressed the sums of non-integer powers of $\mu_i$
directly. Now equation~(\ref{linearcoeff}) reads
\begin{eqnarray*}
\theta_1 = a = \frac{\sum_{i=1}^N (1-\mu_i)(1-I(\mu_i)/I(1))}{\sum_{i=1}^N (1-\mu_i)^2} = \sum_{n=1}^{4}c_n \frac{S_n}{h(N)}
\end{eqnarray*}
and we need  to take the limit as $N\to \infty$ The main challenge for
taking this limit 
is to obtain a closed form expression for the limit of the ratio between the sums of non-integer powers of $\mu_i$ and $h(N)$, terms which 
appear in the ratios $S_1/h(N)$ and $S_3/h(N)$. To evaluate this, we first obtain an expression for the sums of non-integer powers 
of $\mu_i$. In order to do so, we note that, for (integer and non-integer) exponent $k\neq \pm1$,
\begin{eqnarray*}
 \sum_{i=1}^{N} \mu_i^{k} &=& \frac{1}{(N-1)^k}\left(\sum_{j=0}^{N}j^k - N^k\right)\\
                &=& \frac{1}{(N-1)^k}\left(\frac{N^{k+1}}{k+1}-\frac{N^k}{2} + \mathcal{O}(N^{k-1})\right),
\end{eqnarray*}
where in the last step we have used the Euler-Maclaurin summation formula. This implies that
\begin{eqnarray*}
 \frac{\sum_{i=1}^{N} \mu_i^{k}}{h(N)} = \frac{6(N-1)^{(1-k)}}{(2N-1)}\left(\frac{N^{k}}{k+1}-\frac{N^{k-1}}{2} + \mathcal{O}(N^{k-2})\right),
\end{eqnarray*}
whose limit as $N\to \infty$ is easily found as all terms to the right
of the first term in this expression vanish in that limit, 
i.e.,
\begin{eqnarray*}
\lim_{N\to \infty}{\frac{\sum_{i=1}^{N} \mu_i^{k}}{h(N)}} = \lim_{N\to \infty}{\frac{6(N-1)^{(1-k)}}{(2N-1)}\left(\frac{N^{k}}{k+1}\right)} =\frac{3}{1+k}.
\end{eqnarray*}
Using this result, we finally find
\begin{eqnarray*}
\lim_{N\to\infty}{a}=\lim_{N\to \infty}{\sum_{n=1}^{4}c_n\frac{S_n}{h(N)}} = \frac{7}{10}c_1 + c_2 +\frac{81}{70}c_3 + \frac{5}{4}c_4.
\end{eqnarray*} 
\subsection{Limiting coefficients $u_1$ and $u_2$ for the quadratic law when sampling from the non-linear law}
Following the results in Appendix~A, in the general case of all the two-parameter limb-darkening laws 
there are two equations for the parameters $\theta_1$ and $\theta_2$, with
\begin{eqnarray}
\label{twoparcoeff1} \theta_1 = \frac{\beta_2 \alpha_{2,1}-\beta_1\alpha_{2,2}}{\alpha_{1,2}\alpha_{2,1}-\alpha_{1,1}\alpha_{2,2}},\\
\label{twoparcoeff2} \theta_2 = \frac{\beta_1 \alpha_{1,2}-\beta_2\alpha_{1,1}}{\alpha_{1,2}\alpha_{2,1}-\alpha_{1,1}\alpha_{2,2}},
\end{eqnarray}
Specializing to the case of the quadratic law (i.e. $\theta_1=u_1$ and $\theta_2=u_2$), we obtain
\begin{eqnarray*}
\alpha_{n,k} &=& \sum_{i=1}^N (1-\mu_i)^{n+k},\\
\beta_k &=& \sum_{i=1}^N (1-\mu_i)^k(1-I(\mu_i)/I(1)),
\end{eqnarray*}
for $n=1,2$ and $k=1,2$. We note that in this case $\alpha_{n,k}=\alpha_{k,n}$, with
\begin{eqnarray*}
\alpha_{1,1} &=& \sum_{i=1}^N (1-\mu_i)^{2} = \frac{2N^2-N}{6(N-1)},\\
\alpha_{1,2} &=& \sum_{i=1}^N (1-\mu_i)^{3} = \frac{N^2}{4(N-1)},\\
\alpha_{2,2} &=& \sum_{i=1}^N (1-\mu_i)^{4} = \frac{N(2N-1)(3N^2-3N-1)}{30(N-1)^3},\\
\end{eqnarray*}
and the $\beta_k$ are of the form
\begin{eqnarray*}
\beta_1 &=& \sum^{4}_{n=1} c_n S_n,\\
\beta_2 &=& \sum^{4}_{n=1} c_n B_n,
\end{eqnarray*}
with the $S_n$ given in the past sub-section and,
\begin{eqnarray*}
B_1 &=& \frac{2N^2-N}{6(N-1)} + \sum^N_{i=1} 2\mu_i^{3/2}-\mu_i^{5/2}-\mu_i^{1/2},\\
B_2 &=& \frac{N^2}{4(N-1)},\\
B_3 &=& \frac{2N^2-N}{6(N-1)}+\sum^N_{i=1} 2\mu_i^{5/2}-\mu_i^{7/2}-\mu_i^{3/2},\\
B_4 &=& \frac{9N^4-21N^3+14N^2-N}{30(N-1)^3}.
\end{eqnarray*}
The denominator of the expressions for $u_1$ and $u_2$, i.e., the denominator of 
equations (\ref{twoparcoeff1}) and (\ref{twoparcoeff2}), is given by
\begin{eqnarray*}
\alpha=\alpha_{1,2}^2-\alpha_{1,1}\alpha_{2,2} = \frac{(3N^2-3N+2)(N+1)(2-N)N^2}{720(N-1)^4}.
\end{eqnarray*}
We now note that the limits we want to obtain can be written as
\begin{eqnarray*}
\lim_{N\to \infty}{u_1} &=& \lim_{N\to \infty}{\frac{\beta_2 \alpha_{2,1}}{\alpha}}-
\lim_{N\to \infty}{\frac{\beta_1 \alpha_{2,2}}{\alpha}},\\
\lim_{N\to \infty}{u_2} &=& \lim_{N\to \infty}{\frac{\beta_1 \alpha_{1,2}}{\alpha}}-
\lim_{N\to \infty}{\frac{\beta_2 \alpha_{1,1}}{\alpha}},
\end{eqnarray*}
which, using the same methods as in Appendix~B1  gives
\begin{eqnarray*}
\lim_{N\to \infty}{u_1} &=& \frac{12}{35}c_1 + c_2 + \frac{164}{105}c_3 +2c_4\\
\lim_{N\to \infty}{u_2} &=& \frac{10}{21}c_1 - \frac{34}{63}c_3 - c_4
\end{eqnarray*}

\section{Fitting and sampling from skew-normal distributions given
  parameter estimates with asymmetrical errorbars}

Given an estimate of a parameter $\theta$ in the form
$\hat{\theta}^{\sigma_1}_{-\sigma_2}$, where in general $\sigma_1\neq
\sigma_2$, we want to sample points from the posterior distribution of
$\theta$, given the data, only knowing that the distribution is
asymmetric. One choice for performing such sampling is to assume that
the distribution of $\theta$ is a skew-normal distribution with
parameters $\mu, \sigma$ and $\alpha$, which is given by
\begin{eqnarray*}
 p(\theta|\mu,\sigma,\alpha) = p_\alpha\left(\frac{\theta-\mu}{\sigma}\right) \frac{1}{\sigma},
\end{eqnarray*}
where,
\begin{eqnarray*}
p_\alpha(y) = 2\phi(y) \Phi(\alpha y),
\end{eqnarray*}
and,
\begin{eqnarray*}
 \phi(y) = \exp(-y^2/2)/\sqrt{2\pi},\ \Phi(\alpha y) = \int_{-\infty}^{\alpha y} \phi(t)dt.
\end{eqnarray*}
In this distribution, $\mu$ is the mean, $\sigma^2$ its variance and
$\alpha$, also called the shape parameter, defines its skewness. Note
that when $\alpha = 0$, we recover a normal distribution of mean $\mu$
and variance $\sigma^2$.

\subsection{Fitting a skew-normal distribution to observed estimates}

A simple way to obtain the parameters $\{\mu, \sigma, \alpha\}$ of the
distribution for each parameter is to assume that
$\hat{\theta}-\sigma_2$, $\hat{\theta}$ and $\hat{\theta}+\sigma_1$
define the $0.16$, $0.5$ and $0.84$ quantiles of the parameter
distribution (i.e., the quoted value of the parameter defines the
median and the errors define the $68\%$ credibility bands around
it). We can now easily formulate the problem as a non-linear
least-squares problem where the independent, observed, variables are
the observed values at the given quantiles (i.e., $\vec{x} = \{
\hat{\theta}-\sigma_2, \hat{\theta},\hat{\theta}+\sigma_1\}$) and the
dependent variables are the quantiles (i.e., $\vec{y}=\{0.16,0.5,0.84
\}$). Then, we minimize
\begin{eqnarray*}
r = ||\vec{y} - \vec{m}(\vec{x}, \mu, \sigma, \alpha)||,
\end{eqnarray*}
where the $i$-th element of $\vec{m}$ is given by
\begin{eqnarray*}
 m_i = \int_{-\infty}^{x_i} p(\theta|\mu,\sigma,\alpha)d\theta,
\end{eqnarray*}
which can be solved with any non-linear least-squares algorithm such
as Levenberg-Marquardt.

\subsection{Sampling from a skew-normal with known parameters}

Once the parameters for a given skew-normal distribution are known, we
want to sample values from it in a simple and efficient fashion. The
following algorithm has been published by A. Azzalini in his personal
webpage\footnote{\texttt{http://azzalini.stat.unipd.it/SN/faq-r.html}},
but we quote it here for completeness.

First, one has to compute the parameter $\delta =
\alpha/\sqrt{1+\alpha^2}$. With this, one now samples the random
variables $u_0, v$, both of which are independent and have standard
normal distributions, and generate the random variable $u_1 = \delta
u_0 + \sqrt{1-\delta^2}v$ which has correlation $\delta$ with
$u_0$. Then, sample the random variable $z$ which equals $u_1$ if
$u_0>0$ and $-u_1$ otherwise. With this, $z$ has a skew-normal
distribution with zero mean, $\sigma=1$ and shape parameter
$\alpha$. Finally, the random variable $SN=\mu + \sigma z$ has a
skew-normal distribution with mean $\mu$, variance $\sigma^2$ and
shape parameter $\alpha$.

\bsp

\label{lastpage}

\end{document}